\documentclass[namedreferences]{solarphysics}
\usepackage[hyperref,optionalrh,solaromanenum]{spr-sola-addons} 
\usepackage[pdftex]{graphicx}      
\usepackage{color}                 

\chardef\us=`\_

\begin{document}
\begin{article}

\begin{opening}

\title{The new Sunspot Number: assembling all corrections}

\author[addressref={afil1},corref,email={}]{\inits{F.}\fnm{Fr\'{e}d\'{e}ric}~\lnm{Clette}}
\author[addressref={afil1},corref,email={}]{\inits{L.}\fnm{Laure}~\lnm{Lef\`{e}vre}}
 
\runningauthor{Clette et al.}
\runningtitle{The 1981--2015 Brussels--Locarno Sunspot Number}

\address[id={afil1}]{Royal Observatory of Belgium, 3 avenue Circulaire, 1180 Bruxelles, Belgium}

\begin{abstract}
The Sunspot Number, created by R.\,Wolf in 1849, provides a direct long-term record of solar activity from 1700 to the present. In spite of its central role in multiple studies of the solar dynamo and of the past Sun-Earth relations, it was never submitted to a global critical revision. However, various discrepancies with other solar indices recently motivated a full re-calibration of this series.

Based on various diagnostics and corrections established in the framework of several Sunspot Number Workshops and described in \citet{Clette_etal_2014}, we assembled all corrections in order to produce a new standard version of this reference time series. In this paper, we explain the three main corrections and the criteria used to choose a final optimal version of each correction factor or function, given the available information and published analyses. We then discuss the good agreement obtained with the Group sunspot Number derived from a recent reconstruction. Among the implications emerging from this re-calibrated series, we also discuss the absence of a rising secular trend in the newly-determined solar cycle amplitudes, also in relation with contradictory indications derived from cosmogenic radionuclides. 

As conclusion, we introduce the new version management scheme now implemented at the World Data Center - SILSO, which reflects a major conceptual transition: beyond the re-scaled numbers, this first revision of the Sunspot Number also transforms the former locked data archive into a living observational series open to future improvements.   
\end{abstract}


\end{opening}


\section{Introduction} \label{S-Intro} 
Since September 2011, a major end-to-end revision of the sunspot number series has been undertaken in the framework of four successive Sunspot Number workshops held alternately in Europe and the USA \citep{Cliver_etal_2013_CEAB, Cliver_etal_2015_CEAB}. This joint effort involved various scientists working on different issues and epochs in this long series. It was primarily motivated by the large mismatch between the only two long-term sunspot records available thus far: the official Z\"urich-Brussels Sunspot Number (hereafter SN) currently maintained by the World Data Center SILSO \citep{Clette_etal_2007} and the Group sunspot Number (hereafter GN) created more recently by \citet{Hoyt-Schatten_1998a,Hoyt-Schatten_1998b} and ending in 1995. Indeed, while those two references are based on the count of the same base solar features, sunspots, and are thus expected to vary in a very similar way in the course of past solar cycles, the existing series disagreed by more than 40\%, strongly hinting at artificial inhomogeneities in the calibration of one of the series or both.

This work led to the identification of various flaws affecting both series over different time intervals. The diagnostics and the resulting corrections were described in detail in a recent review paper \citep{Clette_etal_2014}. A final important step was accomplished over the past few months (end of 2014 and spring of 2015): all corrections, which were so far identified separately in the original time series, were assembled to derive a new official version of the Sunspot Number. This is thus when the compatibility between different corrections has to be verified and a best estimate of the corrections must be chosen given the uncertainties or the existence of different non-matching determinations. In this paper, we describe this final process that led to the new Sunspot Number series officially released on July $1^{st}$, 2015 by the World Data Center SILSO. This text thus primarily documents the exact modifications included in the new archived series, compared to the original sunspot number that was in use without any modification since a last local adjustment of cycle 5 made by A. Wolfer in 1902 \citep{Wolfer_1902}.

After presenting new conventions adopted together with the new re-calibrated series, we first summarize the correction of variable drifts affecting the ``Brussels'' sunspot number after 1980, which are explained in detail in a companion paper in this issue \citep[hereafter Paper 1]{Clette_etal_2015a}.  Working backwards in time, we then consider the upward bias associated with the introduction of a weighting of spots according to size at Z\"urich in 1947, which affects all values after that year and thus also the Locarno-based sunspot numbers of the ``Brussels'' period. Finally, we describe a third and more local adjustment affecting the first years of the series compiled by Rudolf Wolf starting in 1849.

A corresponding correction and reconstruction of the Group Sunspot Number was also completed by July 2015. We don't describe it here, in particular as this recalibration was done completely independently, using different methods and relying on different underlying observations. The main new version currently available is the Group Number series built using so-called ``backbone'' observers by L. Svalgaard \citep[in this volume]{Svalgaard_2013_wiki, Svalgaard-Schatten_2015}. Here, it is used only for some comparisons. We thus stress here that the corrections to the long-term scale of the new SN series do not rest on any dependency to the GN series. We cared to only use original visual spot and group counts, according to the standard definition of the Wolf number:
\begin{equation}
W_S = k \times (10\,N_g + N_s)    \label{EQ_wolfnum}
\end{equation}
where $N_g$ is the number of sunspot groups, $N_s$ the total number of spots and  $k$ a scaling coefficient specific to each observer.
Therefore, the final comparison between the new SN and GN series discussed in our concluding section gives a meaningful indication of the validity of our corrections and of the improved agreement actually obtained between the two time series. There is no question here of an {\it ad hoc} tweaking of one series to better match the other one, even though the ultimate success in better reconciling the series might inspire this impression, as we observed from the initial concerns expressed by some colleagues.

\section{New conventions} \label{S-Conventions}
As the recalibration described in the following sections leads to rather large changes in the values of the SN, we considered that it was the right time to reconsider some conventions that have been maintained over the years mainly by a passive heritage from choices made long ago. In order to avoid the confusion that would result from successive small changes, we wanted to combine all necessary changes into a single transition.

\subsection{Suppression of the 0.6 Wolf -- Wolfer conversion factor}
The main change is the suppression of the conventional 0.6 factor introduced by A.\,Wolfer in 1894 and applied to all SN values after that year. The value of this conversion factor results from 17 years of parallel observations (1877 -- 1893) done by R.\,Wolf and his younger assistant A.\,Wolfer. It scales the higher values obtained by Wolfer using the standard 83\,mm Fraunhofer refractor and modern counting rules down to the lower numbers obtained by Wolf using mainly smaller portable refractors and neglecting the smallest short-lived spots in order to be more consistent with ancient historical observers. Quite naturally, when Wolfer took over as Director of the Z\"urich Observatory after Wolf's death, he chose to scale down his daily observations by the 0.6 factor, in order to align them on the existing long series built manually by Wolf over several decades. Since then, this tradition was just continued up to these days.

However, now in 2015, we have accumulated more than 120 years of modern counts, i.e. much more than the 44 years of Wolf's own observations. Moreover, with modern computers, rescaling thousands of past values does not pose a challenge like at Wolfer's epoch. Therefore, continuing this rescaling does not makes sense anymore. Still, the strongest motivation is in fact the need to eliminate the enduring confusion that this down-scaling causes nowadays to the scientific community and SN users at large. Indeed, without insight in the historical construction of the SN, our users don't understand why the official SN is always much lower than other more recent sunspot counts (e.g. the NOAA and Boulder numbers) or lower than what they can actually count themselves with their own instrument. 

Consequently, we decided to eliminate this factor. When constructing the new SN series, the original series was thus first rescaled by dividing all values by the constant factor 0.6 before applying the various corrections. This simply means that Wolf numbers from Wolfer define the scale of the entire series in place of Wolf. The conversion is thus equivalent to a change of unit. Adopting Wolfer as new reference also makes sense when considering the history and properties of the SN series. Indeed, all available publications indicate that Wolfer used exactly the same counting method and the same instrument for all his observations. Therefore, the period 1894 -- 1926 is the most coherent in the SN series. By contrast, we know that in the course of his career, Wolf used several instruments at different times and he only progressively established and refined the determination of k personal coefficients \citep{Kopecky_etal_1980, Clette_etal_2014}. Likewise, after Wolfer, changes in the counting methods caused jumps and drifts in the SN scale as described hereafter. 

\subsection{Elimination of the 0 -- 11 quantification.} 
When it was produced by the Observatory of Z{\"u}rich from 1849 to 1980, the SN was simply the Wolf number of the Z{\"u}rich observer on most days. Auxiliary stations were only used for filling the gaps, on days when no observation could be made in Z{\"u}rich. On such days, the SN was derived by an average of the Wolf numbers of other stations, normalized by their k personal factors. However, at low solar activity, such an average can produce a continuous range of values above 0, while the Wolf Number of a single observer (equation \ref{EQ_wolfnum} without k factor) cannot take a value between 0 (no spot) and 11 (single spot). The adopted practice was thus to simulate a single observer: at the very end of the calculation, if the daily average is non-null, it is set to 11.

When the calculation of the sunspot number was taken over by the Royal Observatory in Brussels, this rule was maintained for consistency with the earlier Z{\"u}rich data. However, as the new method included all stations in the calculations of all days, this rule is applied more frequently than before. Still, as it applies only for very low levels of solar activity, it is only playing a role during short periods around the times of cycle minima. Nevertheless, we decided to remove this rule, as it may lead to a slight overestimate of the SN at low activity, since it artificially raises some low values. This 0 -- 11 jump may also contribute to the non-linear relation between the SN and other indices like the $F_{10.7cm}$ radio flux \citep{Holland-Vaughn_1984, Johnson_2011}. We must also point out that accepting continuous values for low SNs is not unprecedented. Over the period when he was observing in parallel with Wolfer, Wolf simply averaged the simultaneous numbers without further modification. As shown by \citet[Fig. 4]{Clette_etal_2014}, this is visible in the histogram of original SN values as the absence of a 0 -- 11 gap over the period 1877 -- 1893. 

The elimination of this 0 -- 11 jump applies both to the new SN values appended each month to the SN series and to the corrected SN values, where the application of a correction factor to the original SN values may produce a result between 0 and 11. In the future, in the framework of a full recalculation of the ``Brussels'' SN from 1980 onwards, we plan to apply it as well. 

\subsection{New symbol $S_N$}
So far, the base symbol for the SN was the letter $R$ adopted by Wolf in his publications where it stands for ``relative''. Indeed, noticing that different observers gave different counts and considering that this index cannot be expressed in a physical unit, he chose to describe the number as relative. Although this is largely correct, direct comparisons with modern measurements of solar properties have now become possible and indicate a very high correlation with quantitative solar parameters, in particular with direct tracers of the magnetic flux emergence at the solar surface: see e.g. \citet[Fig. 6]{Stenflo_2012} for a recent comparison with the global unsigned vertical magnetic flux density. Therefore, considering also the extensive statistics over many observers leading to the current SN, the relative nature of the SN does not appear as a unique characteristic distinguishing it sharply from other more recent solar indices. We also note that the letter R does not evoke anything related to the Sun.

Moreover, over the years, multiple designations have also appeared in the literature to distinguish different periods or different instances: $R$, $R_Z$ (Z{\"u}rich SN before 1980), $R_i$ (International SN after 1980), $R_A$ (American SN). $R$ was also used for proxies of the actual SN, like $R_G$ designating the Group sunspot number. In order to reduce this proliferation, we wanted to replace those symbols by a new one, reducing the highlight put on ``relative''. A new symbol would also help users to directly make a clear distinction between the new SN series and the original one. We settled for the letter S that simply stands for ``sunspot'' and added a subscript N for ``number'', close to the initials and to distinguish it from e.g. A for area, etc. We use the new adopted symbol ``$S_N$'' in the rest of this paper. In the same vein, for coherency, we would suggest ``$G_N$'' as new symbol for the Group Number (in place of $R_G$).

\section{The Locarno drift correction (1981 -- 2015)} \label{S-LocCorr} 
Through a global statistical analysis exploiting the huge data archive from the SILSO network, we have identified a large variable trend in the scale of the SN after 1981 \citep{Clette_etal_2014}, i.e. over the period when the SN was based on a new pilot station, the Specola Solare Observatory in Locarno, Switzerland. This station was chosen as a replacement for the Z{\"u}rich Observatory, when the latter was closed in 1980. As this station had been trained by the Director of the Z{\"u}rich Observatory, Max Waldmeier, and had provided parallel complementary observations to Z{\"u}rich from 1958 to 1980, it offered the best guarantee to ensure a seamless continuity in the scale of the SN across the 1981 transition. As explained in Clette et al. 2015 (Paper 1, in this issue), a good continuity was indeed achieved. However, soon after 1981 and until recent years, the Wolf Numbers from the Specola station started to be overestimated, and later underestimated between 1995 and 2005, due to different local factors that are not fully identified yet.

\begin{figure}[bt] 
	\centerline{\includegraphics[width=1.0\textwidth,clip=true,trim= 15 0 5 0,clip=true]{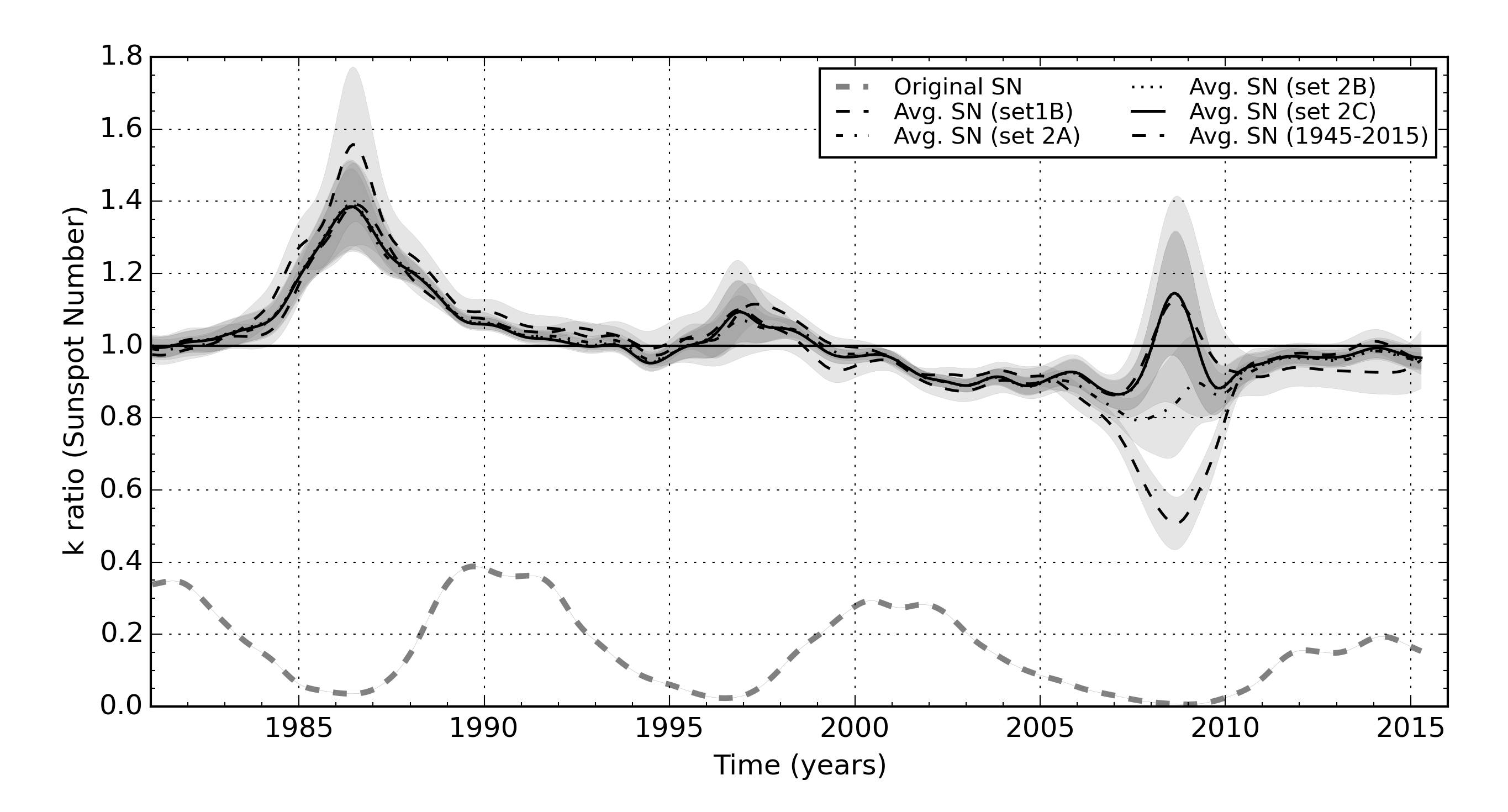}}
	\caption{k ratios between the average SN obtained from different sets of stations and the original SN series. The grey shading gives the standard error, which peaks at the time of solar cycle minima (The evolution of the cycles is given below by the gray dashed curve). The original sunspot number was first overestimated between 1983 and 1994 then underestimated after 2000 before returning close to the 1981 level in recent years. (Figure 6 from paper 1, \cite{Clette_etal_2015a})} 
	\label{Fig-LOdrift}
\end{figure}

As the scale of the SN is tied to the reference provided by the pilot station over time scales longer than one month (the base time interval over which the personal coefficients of individual stations are determined since 1981), it led to the scale variations shown in figure \ref{Fig-LOdrift}, which is taken from \citet{Clette_etal_2015a}. The ratio between the re-constructed network average and the original SN thus provides the correction factor to be applied to the original series to correct for the drifts. As the base time unit for deriving the k coefficient was one month, we thus derived the monthly mean ratios over the entire time interval 1981 -- 2015 and the monthly mean factor for each month was applied to all daily values of the corresponding month, which gives the new daily Sunspot Number series $S_N$.

However, this correction factor is only relative. This is why an extended analysis was also carried out, spanning the interval 1945 -- 2015, thus including more than 30 years of the Z{\"u}rich era 1945 -- 1980, which forms our reference. We established that no systematic trend was present over this time interval and we calibrated the scale of the 1981 -- 2015 numbers versus the preceding period within 1\%. This was obtained by comparing the average ratio between the original SN series and a multi-station average for the interval before 1980 and the equivalent ratio between the SN series multiplied by the variable correction and the same multi-station average after 1980 \citep[Fig. 10]{Clette_etal_2015a}.

This thus connects the ``Brussels'' era of the SN to the preceding Z{\"u}rich scale. However, as we will see in the next section, it turns out that the Z{\"u}rich SN was affected by another bias since 1947. As this was caused by the introduction of a new counting method also used by the Specola station up to the present, this other correction must be combined with the 1981 -- 2015 drift, as it played a role up to the present, including at the time of the 1980 Z{\"u}rich to Brussels transition.

\section{The ``Waldmeier'' jump (1947 -- 1980)} \label{S-WaldCorr} 
An abrupt upward jump in the scale of the SN in 1947 was found by comparisons with various parallel solar data series, including the GN series \citep{Clette_etal_2014, Svalgaard_2015a, Svalgaard_2015a_arXiv}. The cause of this transition was traced to the introduction of a new counting method by Max Waldmeier, the last Director of the Z{\"u}rich Observatory \citep{Clette_etal_2014}. In this method, large spots with an extended penumbra are counted for more than one, in a range from 2 to 5 according to the spot size. More subtle rules include a global estimated count for clusters of small spots within a group (which can lead to a lower count of the smallest spots) and a compensation of the degraded visibility of sunspot groups located near the solar limb. 

This new method thus definitely deviates from the base Wolf Number definition that had been in use until then, and it must lead to a net global increase of the resulting sunspot numbers. Unfortunately, this change was hardly documented by Waldmeier \citep{Waldmeier_1948, Waldmeier_1968, Clette_etal_2014}, and most of what we know about this method comes from the current daily use of weighted counts by the main Locarno observer, Sergio Cortesi, who was personally trained by Waldmeier in 1958. 

Because of this lack of documentation, we also don't know when this new method was actually implemented. Although earlier Z{\"u}rich observations by W. Brunner and his assistants give evidence that weighted counts were already used before 1940, at least occasionally (S.\,Cortesi and M.\,Cagnotti, private communication), the rather sharp jump in 1947 suggests that this year marks the start of its systematic application by Waldmeier and his staff.

\subsection{Contradictory estimates of the ``Waldmeier'' jump}  \label{SS-WaldContradic}
The first evidence for the Waldmeier jump resulted from comparisons between the original SN and various parallel solar series that cannot be affected by the Z{\"u}rich reference: Wolf numbers from individual long-duration observers (e.g. Madrid Observatory), sunspot areas from the Greenwich photo-heliographic catalog, yearly means of the diurnal modulation of the East component of the geomagnetic field, ionospheric soundings \citep{Cliver-Svalgaard_2007_AGU, Svalgaard-Cliver_2007_AGU, Clette_etal_2014}. Most of those comparisons indicate an inflation of the SN of about 20\% after 1947 \citep{Clette_etal_2014}. Of course, as the series used for comparison may be affected by other sources of error, the uncertainty on this determination is rather large and this is confirmed by the wide range of derived values from about 15 to 25\% \citep{Svalgaard-Bertello_2009_SPD, Clette_etal_2014}.

In this respect, significantly lower values, near 12\%, were obtained by Lockwood et al. in two recent analyses resting either on the Greenwich-based GN, on the total spot area \citep{Lockwood_etal_2014} or on the first ionospheric soundings \citep[this volume]{Lockwood_etal_2015a}. As the latter study is very recent, we will not (yet) consider it in detail here. We just note that it involves a data series starting only in 1932, i.e. shortly before the Waldmeier transition, and requires several hypotheses in order to transform raw measurements into an unbiased solar proxy. On the other hand, \citet[hereafter LOB2014]{Lockwood_etal_2014} consider the ratio between the SN and two direct sunspot indices (NB: their use of geomagnetic indices does not lead to a better accuracy than all the above-mentioned comparisons). By repeating their analysis, we found two main flaws that each lead to an underestimate of the 1947 Waldmeier jump.

Considering first the Group Number, LOB2014 use the numbers from the original Greenwich photographic catalog. This series, which forms the base for the original GN series \citep{Hoyt-Schatten_1998a, Hoyt-Schatten_1998b}, is in fact affected by a large upward trend between 1885 and 1915 \citep[this volume]{Cliver_etal_2013_CEAB, Clette_etal_2014, Cliver-Ling_2015}. Therefore, all Greenwich GN values are underestimated by an amount that increases progressively before 1915 and reaches 40\% at the start of the GN series in 1874. We thus expect the SN/GN ratio to be strongly overestimated over a significant part of the 1874 -- 1945 interval preceding the 1947 jump. Likewise, when considering their long series obtained by extending the Greenwich data by SOON GNs after 1976, they take the ratio between the GN and the original uncorrected SN series up to 2013. As explained in the preceding section, the SN suffers from variable trends over this time interval. This may thus also produce a bias in the SN/GN ratio, although in this case, the amplitude of the SN correction is lower and the correction varies both upward and downwards inside the 1945 -- 2013 time-averaging window.

\begin{figure} 
	\centerline{\includegraphics[width=1.0\textwidth,clip=true,trim= 5 0 5 0,clip=true]{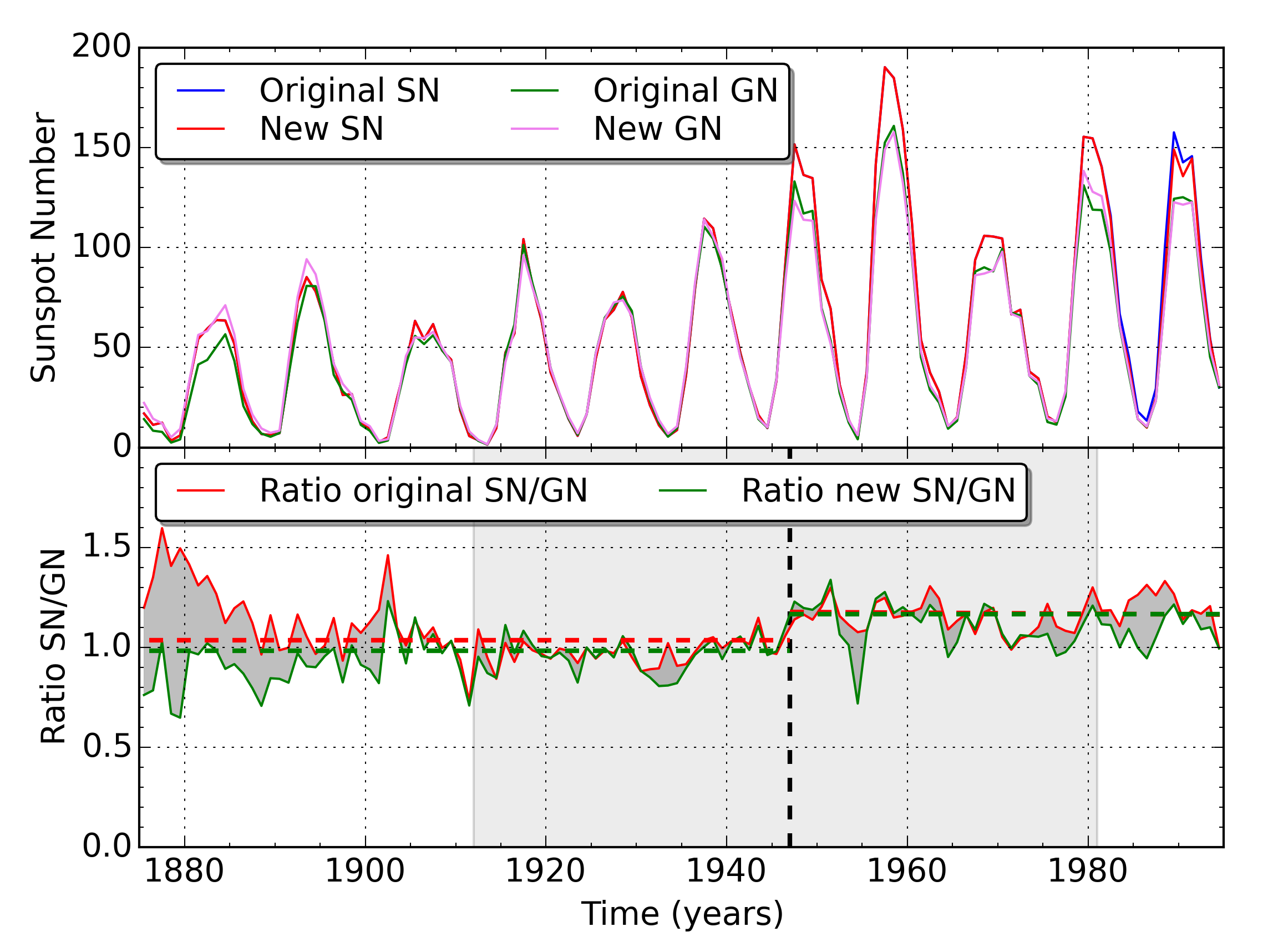}}
	\caption{Comparison of the yearly mean SN/GN ratios (lower panel) for the original SN and GN series (red line) and for the corrected series (green line). The new SN is identical to the original SN except for the correction of the Locarno drift after 1981. The new GN used here is the ``backbone'' GN from \citet{Svalgaard-Schatten_2015}. Both series agree closely over the shaded time interval but diverge significantly before and after it, due to known inhomogeneities in the original series (see main text). The vertical dashed line marks the Waldmeier jump, while the horizontal red and green dashed lines show the average ratios for the original and new series respectively, over the 1872 -- 1946 and 1947 -- 1995 intervals. All four SN and GN series themselves are plotted in the upper panel.} 
	\label{Fig-WaldJumpSNGN}
\end{figure}

We thus need to quantify the biases resulting from the use of original uncorrected series, i.e. induced by the ``crosstalk'' between this correction and other uncorrected inhomogeneities. For this, we repeated the determination of the averages before and after the 1947 transition  but replacing the original GN by the reconstructed ``backbone'' GN from \citet{Svalgaard-Schatten_2015}, normalized to the original GN over the interval 1874 -- 1947. Here, for the original GN, we use the GN series by \citet{Hoyt-Schatten_1998a, Hoyt-Schatten_1998b}, which is directly based on the Greenwich catalog. Likewise, we replaced the original SN series by this same series corrected only for the 1981 -- 2015 Locarno drift (correction normalized to the whole 1945 -- 2015 interval as described above in section \ref{S-LocCorr}). The comparison between the two pairs of series and the two corresponding ratios is shown in figure \ref{Fig-WaldJumpSNGN}.  

The large upward deviation before 1915 can be recognized by the shaded area. Thereafter, the original and new series match closely until 1981, when a smaller discrepancy appears, corresponding to the Locarno drifts affecting the SN series. We then determined the average SN/GN ratios before and after 1947 using 6 different methods as cross-validation, in the same manner as for our earlier 1981 -- 2015 SN reconstructions \citep{Clette_etal_2015a}. We find that the original average before 1947 is 5\% higher than when using corrected series. After 1947, the original average ratio is just slightly higher than with the corrected SN series, by barely 1\%, confirming that for this time period, the upward and downward drifts mostly compensate each other in the global average. Taking the ratio between those two averages to determine the jump amplitude, we thus find that the low ratio of 1.126 derived in LOB2014 was underestimated by 4\%, due entirely to overlooked inhomogeneities in the base data series. A similar result, but with larger uncertainties, is obtained by simply using the original series but restricting the analysis to the homogeneous interval 1915 -- 1980 (shaded interval in figure \ref{Fig-WaldJumpSNGN}). With cleaned series, we thus find a new higher ratio of $1.17 \pm 0.01$, which agrees much better with the other determinations.

Considering now the comparison of the SN series with the Greenwich sunspot areas ($A_G$), we find that the published LOB2014 $f_R$ jump factor for $A_G$ is also affected by the choice of the outer boundaries of the analyzed time interval, but even more by the choice of the splitting year between the two averaging intervals. In order to analyze this effect, we fully replicated the method described by LOB2014. As base data, we derived the group count $N_G$ and total group area $A_G$ from the Greenwich catalog over the interval 1874 -- 1976 
 and we took the original SN series $R_i$. For all series, we computed the yearly means. As the relation between $A_G$ and SN is non-linear, we used $A_G^{\,m}$, with an adjusted m exponent, like in LOB2014 (m stays close to their 0.871 value). The jump factor $f_R$, as defined by LOB2014 was obtained by the minimization of the average residual differences between the SN series and each of the comparison series $N_G$ and $A_G$ over the time intervals before and after the splitting year, which was set in 1945 by LOB2014. 

Here, instead of studying only one single set of time intervals, we varied the starting time from 1874 to 1930, in order to evaluate the possible effect of an early inhomogeneity in the Greenwich group areas and we also varied the splitting time $T_s$ separating the two time intervals over which the mean residuals are derived. 

\begin{figure} 
	\centerline{\includegraphics[width=0.7\textwidth,clip=true,trim= 5 0 5 0,clip=true]{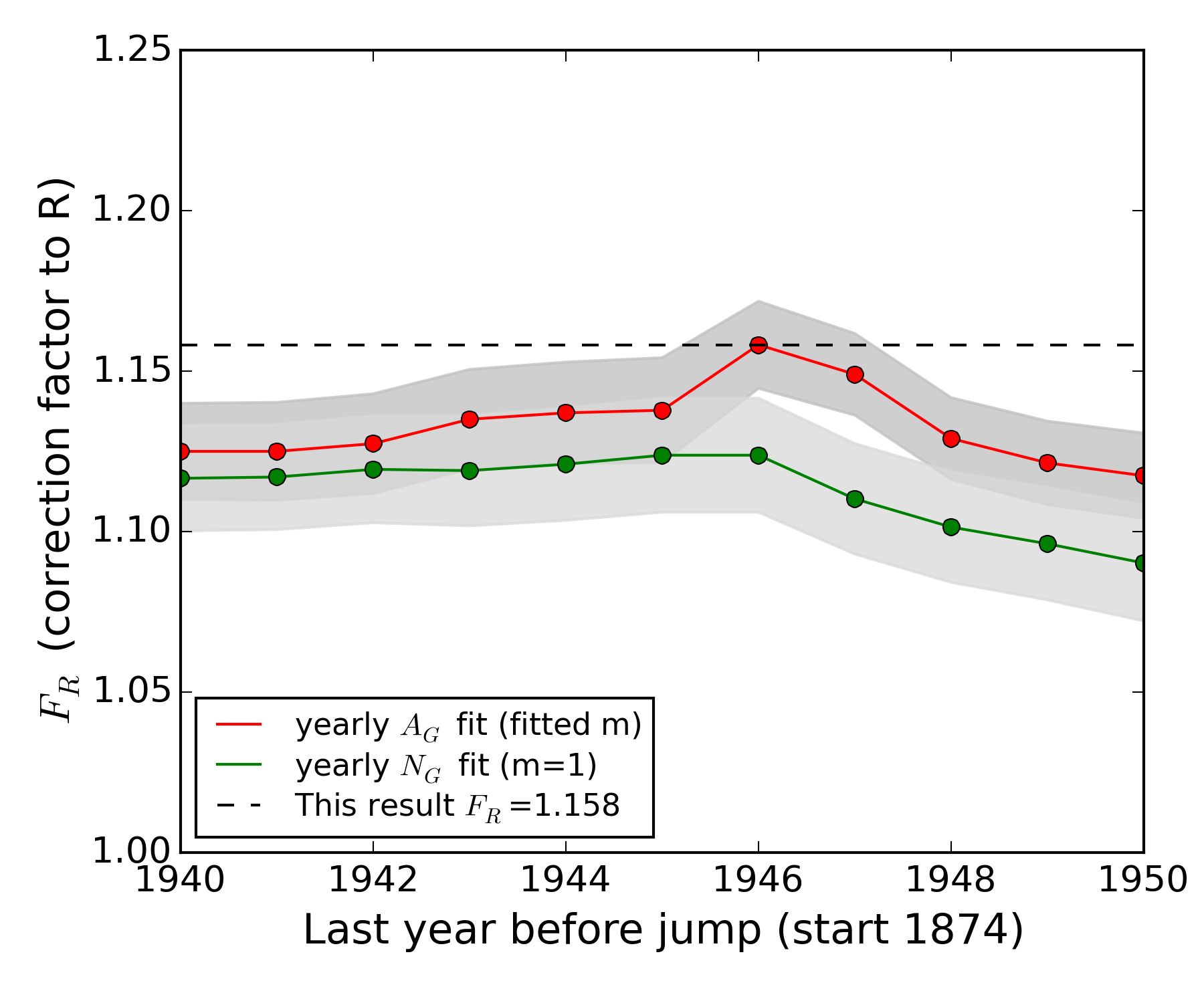}}
	\centerline{\includegraphics[width=0.7\textwidth,clip=true,trim= 5 0 5 0,clip=true]{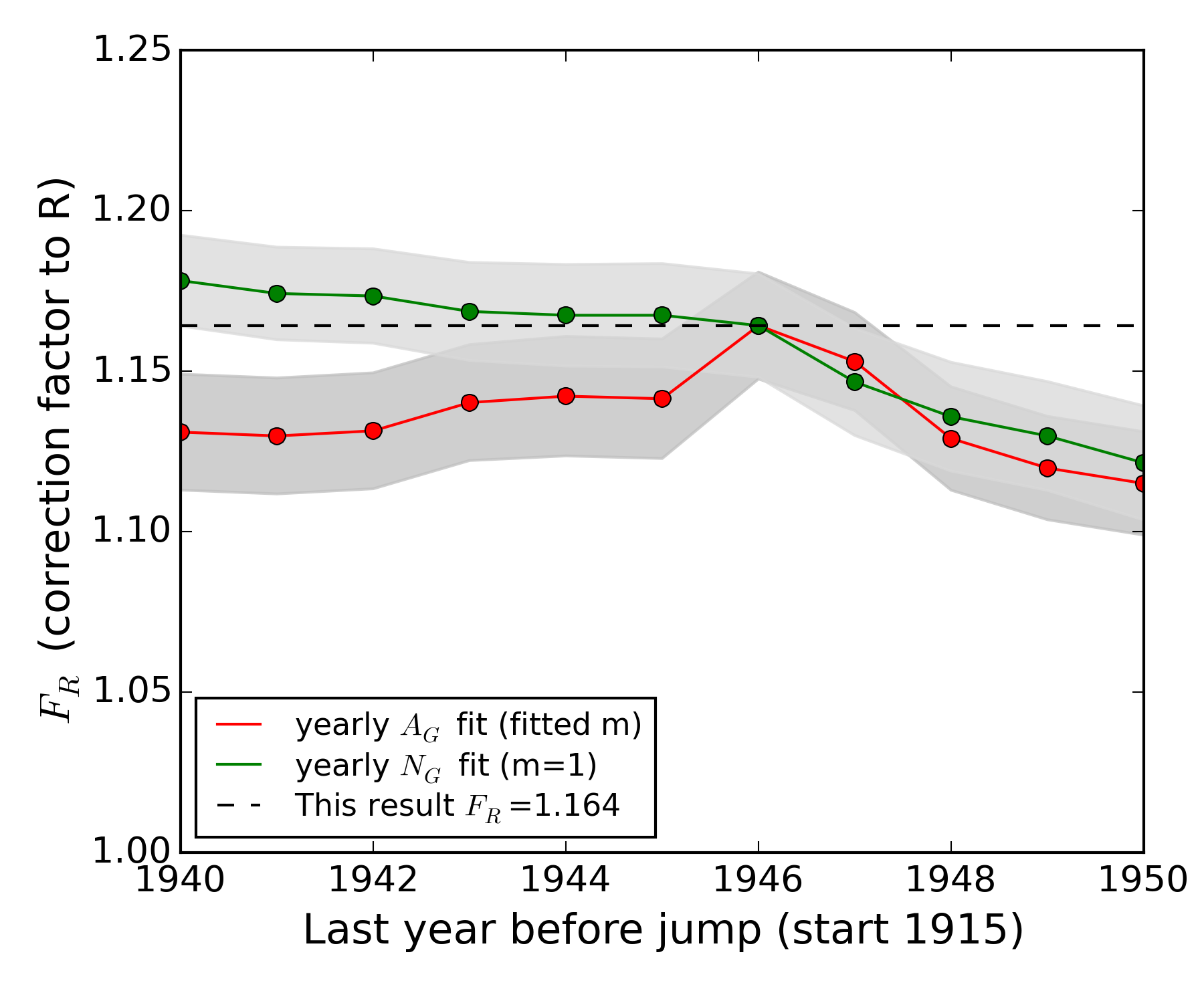}}
	\caption{The Waldmeier jump factor $f_R$ as a function of the splitting year $T_s$ chosen to separate data before and after the transition, for an analysis interval starting in 1874 (top) and in 1915 (bottom). The green curves correspond to the $f_R$ factor for the SN/$N_G$ ratio, while the red curves correspond to $f_R$ for the SN/$A_G$ ratio. The grey shading indicates the $1 \sigma$ uncertainty on $f_R$. A transition or local maximum appears in both curves for 1946, which matches the actual time when the Waldmeier method was applied systematically. By excluding the time interval 1874 -- 1915 (bottom plot), both factors increase, though only slightly for the SN/$A_G$ ratio. Both factors only match in the lower plot for $T_s = 1946$, which marks the maximum of $f_R$ for the SN/$A_G$ ratio.} 
	\label{Fig-WaldJumpSNAs}
\end{figure}

Figure \ref{Fig-WaldJumpSNAs} illustrates the results in two plots showing the variation of the jump factor $f_R$ as a function of the splitting year $T_s$, for two choices of the starting year (1874 or 1915) of the analysis interval. Both plots show a local maximum for the SN/$A_G$ ratio and an inflexion in the SN/$N_G$ ratio in 1946. This feature is expected if the SN series contains a sharp jump in its average scale. Indeed, if the splitting year does not match the true time of the jump, the latter falls inside one of the two averaging intervals: e.g. a few years with the lower scale will be incorrectly included on the side where values have a higher scale, thus reducing the corresponding average below its actual value. This effect is clearly seen in the SN/$A_G$ ratio, which peaks in 1946, confirming that the actual Waldmeier transition occurred between 1946 and 1947, as already found by \citet{Clette_etal_2014}. However, LOB2014 incorrectly chose 1945. While we find the same value of the $f_R$ ratio ($\approx 1.14$) as in the LOB2014 SN/$A_G$ study for $T_s = 1945$ and the same starting time in 1874 (Fig. \ref{Fig-WaldJumpSNAs}, top), we can see that it underestimates the actual value of 1.158 found when splitting in 1946.

Comparing now the analyses starting in 1874 and 1915 (two plots in Fig. \ref{Fig-WaldJumpSNAs}), we can see that the SN/$N_G$ ratio increases from 1.124 to 1.164 (with $T_s = 1946$), confirming the 4\% underestimate caused by the inhomogeneity in the Greenwich group counts before 1915 (see above and figure \ref{Fig-WaldJumpSNGN}). The SN/$A_G$ ratio also shows an equivalent increase when avoiding this early part of the Greenwich catalog, but only from 1.158 to 1.164, i.e. hardly 0.5\%. The inhomogeneity is thus clearly much smaller for this quantity in the Greenwich data. Finally, the $f_R$ ratios for $N_G$ and $A_G$ come in full agreement only when excluding the pre-1915 part of the Greenwich data (Fig. \ref{Fig-WaldJumpSNAs}, bottom plot) and adopting 1946 as splitting year, with $f_R = 1.642 \pm 0.023$. We can thus conclude that the abnormally low jump factors found by LOB2014 are due primarily to the 1874 -- 1915 Greenwich trend for what concerns the SN/$N_G$ ratio. On the other hand, they are due primarily to an improper choice of 1945 as splitting time for the SN/$A_G$ ratio. 

Finally, combining the $\sigma$ estimation from the LOB2014 method with the uncertainty caused by different other factors (variations with $T_s$ or start time), we find that the total uncertainty on the $f_R$ ratio should range between 3 and 3.5\%, i.e. much larger than the $\approx 1.8\%$ value given by LOB2014, which thus seems to be underestimated. This may also be associated with a partial correlation between numbers over successive years, a possibility which requires further investigation.

Although we could reconcile this recent determination with other comparisons with parallel solar indices, we must conclude that the reliability of such comparisons with external data is inevitably reduced by the presence of other disturbing factors and defects in those parallel series, unrelated to the SN itself. Therefore, a more direct approach is definitely needed to quantify the inflation of the SN produced by the sunspot weighting. 

\subsection{A direct determination of the inflation factor}
The most direct way to quantify the effect of the sunspot weighting is by comparing simultaneous counts made by the same observer but according to the two different methods. This idea was implemented by Svalgaard \citep{Clette_etal_2014, Svalgaard_2015a, Svalgaard_2015a_arXiv} who re-counted the spots and groups from past Locarno sunspot drawings and in addition, prompted the current prime Specola observer, Marco Cagnotti, to do actual double counts as part of the current observing routine \citep[Fig. 43]{Clette_etal_2014}. Past analyses of those data indicated a mean inflation factors of 1.165 to 1.18, but also showed that the inflation factor varies with the sunspot number, i.e. with the level of solar activity \citep[Fig. 44]{Clette_etal_2014}. Such a variation naturally results from the fact that at high activity levels, a large fraction of all spots are well developed with extended penumbrae, while near cycle minima, small A and B-type sunspot groups dominate, thus with most spots counted as 1 in the weighted scheme just like with the normal Wolf formula.

For a final determination of the sunspot weighting correction, we thus started from the data set collected by Svalgaard. The drawings that were recounted at this stage cover the following periods: the year 1997 and 2003 to 2014 (total of 3661 daily observations). The counts do not involve any interpretation: they only consist in counting the number of groups and spots drawn on the sheet by the Locarno observer and applying the standard Wolf number definition (\ref{EQ_wolfnum}). The weighted sunspot numbers are simply the Wolf numbers reported to the WDC -- SILSO for the same dates. However, at the Specola observatory, the drawings are produced separately from the standard counts made for the determination of the SN. While the latter are done at the eyepiece (aerial image) with the telescope aperture stopped down at 83\,mm (identical to the standard Z{\"u}rich refractor), the drawings are made by projection on a paper sheet (25\,cm diameter image) and with the full 150\,mm aperture of their Zeiss coud{\'e} refractor. Occasionally, the time of the two observations may also be significantly different. Therefore, the number of spots in the drawings do not match necessarily exactly the eyepiece counts. 
  
\begin{figure} 
	\centerline{\includegraphics[width=0.7\textwidth,clip=true,trim= 5 0 5 0,clip=true]{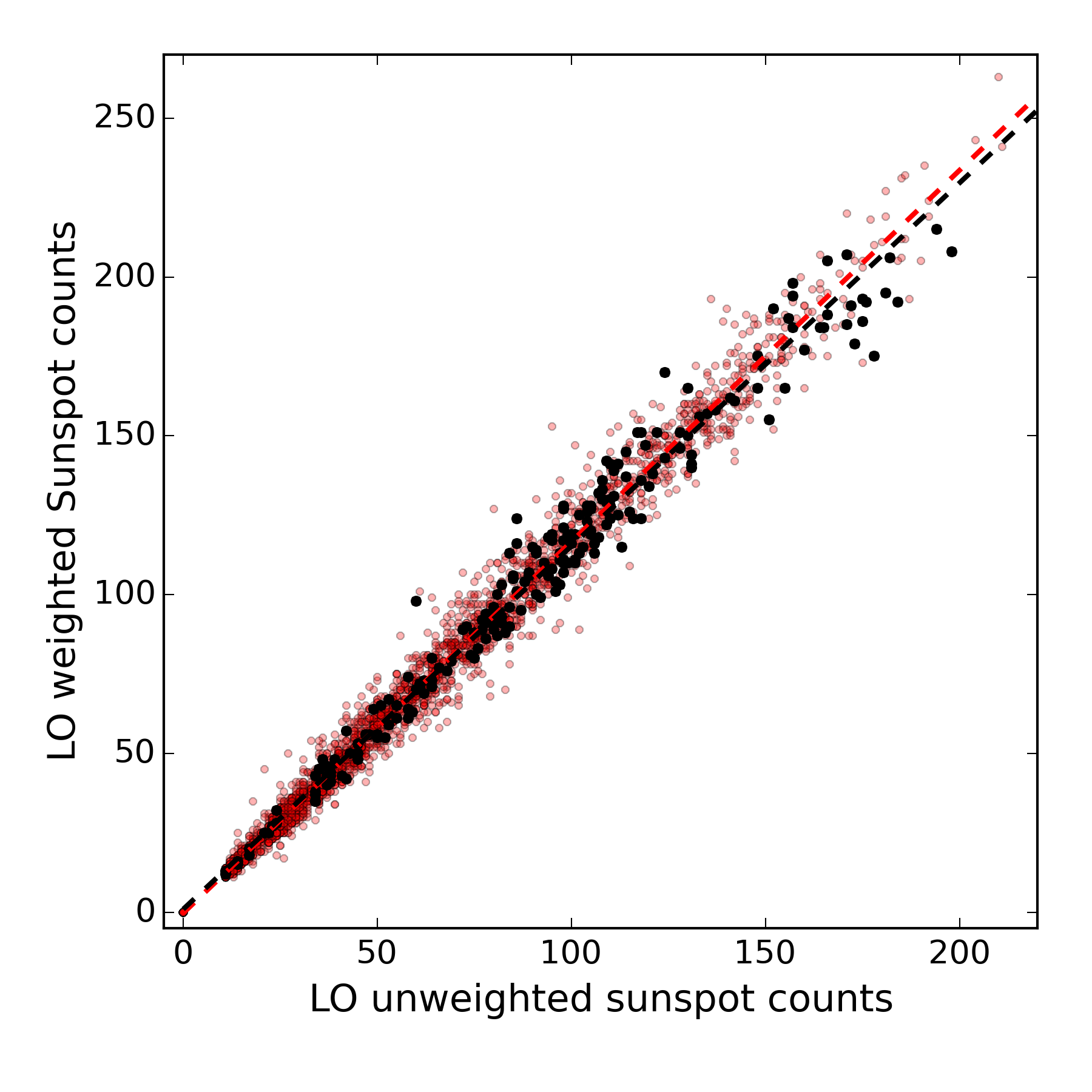}}
	\caption{Scatterplot of the simultaneous weighted versus unweighted counts for two different data sets used in this study: the unweighted counts obtained from the Specola-Locarno drawing archive (red dots) and those obtained simultaneous with weighted counts by Marco Cagnotti at this same observatory over 2014 -- 2015 (black dots). The weighted counts are those provided by the Specola station to the WDC - SILSO. The two dashed lines are linear regressions on both data sets (with corresponding colors), which prove to have almost identical properties.} 
	\label{Fig-SvalVSCagno}
\end{figure}

In order to validate this data set, we also used the direct double counts provided by the Specola Observatory (main observer: M.\,Cagnotti, with some observations by S.\,Cortesi and other alternate observers) and stored in the WDC -- SILSO database for the last two years, from August 2014 to June 2015 (215 daily observations). Here the number of data points is more limited, as well as the range of SN values observed over this recent period, but in this case, we compare  two observations made with rigorously identical setups (eyepiece counts at 83\,mm aperture), by the same observer. The only difference between the two sunspot numbers is the counting method. In figure \ref{Fig-SvalVSCagno}, we over-plot the two sets of daily pairs of numbers, in standard versus weighted number coordinates, showing that both sets perfectly coincide. As the points are largely distributed along a line, we applied a linear regression to both sets, which gives respectively:
\begin{equation}
S_W= - 0.24  (\pm 0.25) + 1.169 (\pm 0.004) S_U    \label{EQ-RegWUDrawings}
\end{equation}
for the drawing set, and 
\begin{equation}
S_W= 1.00 (\pm 2.25) + 1.14 (\pm 0.03) S_U    \label{EQ-RegWULocarno}
\end{equation}
for the eyepiece counts, with $S_W$ and $S_U$ designating respectively the weighted and unweighted SN.
 
Therefore, we can conclude that both sets agree within the uncertainties, and no systematic difference can be found between the counts derived from the drawings and the corresponding eyepiece counts.

Following this verification, we apply our analysis to the larger data set based on re-counted drawings, as it also spans a longer duration, including sections of past solar cycles that reached larger values of the SN number. However, as earlier cycles reached higher peak values of the SN, in particular cycles 18 and 19 that immediately followed the 1947 Waldmeier transition, the relation we can derive from this more recent period will need to be extrapolated above the observed range of SN values, which requires particular care.

Working on yearly means of the original SN and of the inflation factor W derived from the double counts, Svalgaard had derived a preliminary linear relation published in \citet{Clette_etal_2014}: 
\begin{equation}
W= 1.123 (\pm 0.006) + R_i / 1416 (\pm 140)       \label{EQ-WregSvalg}
\end{equation}
with $R_i$ the original SN, which includes the sunspot weighting effect.

\begin{figure} 
	\centerline{\includegraphics[width=0.85\textwidth,clip=true,trim= 5 0 5 0,clip=true]{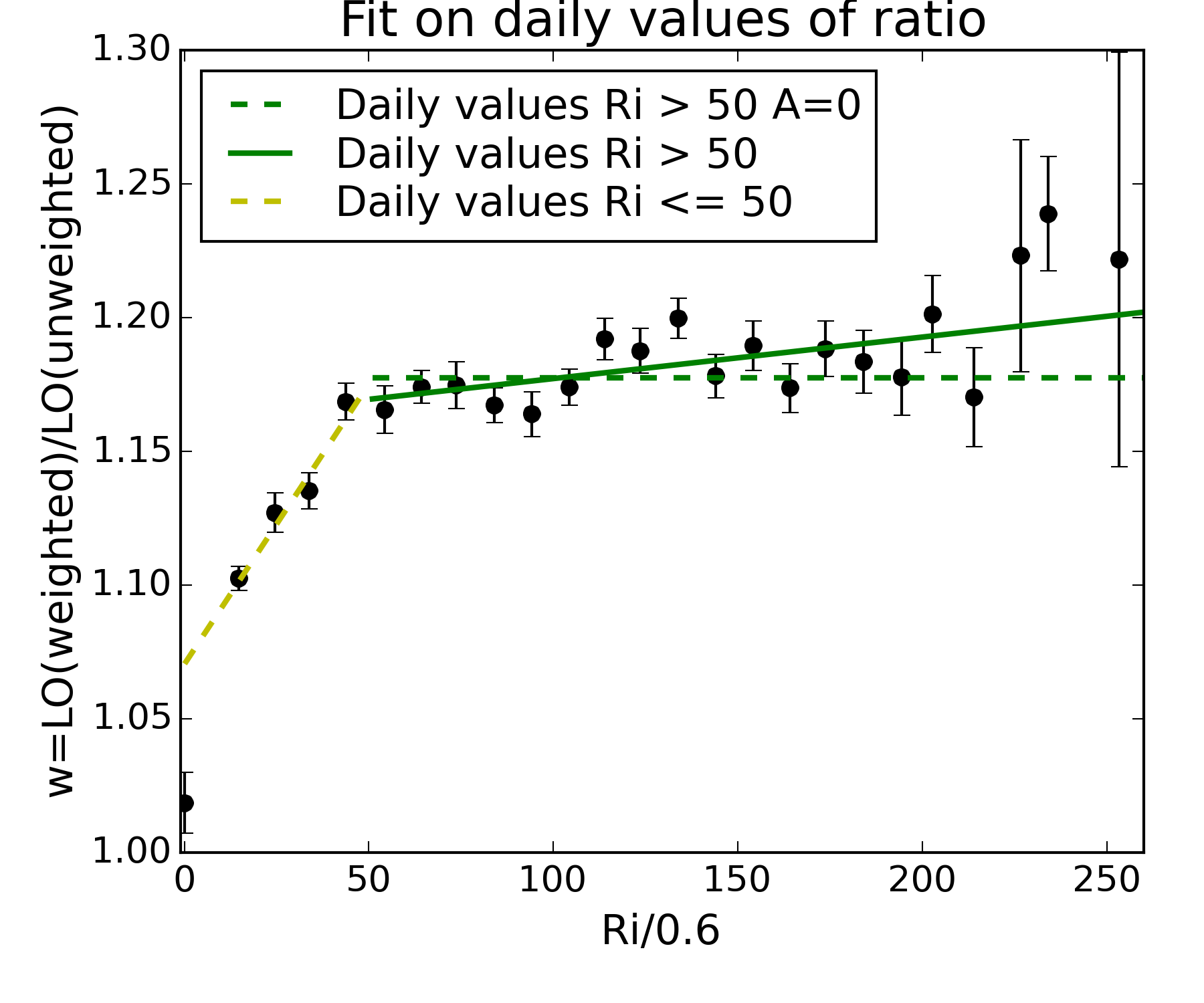}}
	\caption{Weighting inflation factor W as a function of the original SN $R_i^*$ (which includes the weighting effect) scaled without the 0.6 Z\"urich factor. Daily values were averaged over equal bins spanning 10 SN units. The standard error of the means are given by the error bars. The brown dashed line is a linear fit to the values below $R_i^* = 50$. The green solid line is the corresponding fit for $R_i^* > 50$, while the green dashed line simply gives the average W factor over that same interval.} 
	\label{Fig-DailyWvsSN}
\end{figure}

This relation confirmed the expected increase of the inflation factor W with the SN value but it also permitted high values of the W factor ($> 1.25$) for the highest cycles, i.e. well above several estimates coming from comparison with parallel data series. In order to better establish the relation between W and the SN, we consider here the individual daily pairs of values, without any time averaging. Figure \ref{Fig-DailyWvsSN} shows the resulting distribution of W as a function of $R_i^* = R_i/0.6$ (the original SN series divided here by 0.6 in accordance to the new scaling convention for the entire SN series. See Section \ref{S-Conventions}). Daily W factors show a very large dispersion. However, as we collected a very large number of values, for clarity, we binned the values according to the $R_i^*$ value, in intervals of 10 SN units, leading to rather precise mean values. The standard error is typically between 1 and 2\%, except for the highest $R_i^*$ for which the number of available values drops steeply.

Those mean points show an initial steep rise from about 1 to 1.18 for low SN values up to $R_i^* = 50$. The linear fit over that range leads to: 
\begin{equation}
W= 1.07 (\pm 0.02) + 0.0021 (\pm 0.0006) R_i^*       \label{EQ-Wreg<50}
\end{equation}

Above this limit, the W factor abruptly ceases to increase and remains almost constant with a only a slight upward slope. \citet{Svalgaard_2015a, Svalgaard_2015a_arXiv} finds very similar results but still keeps a moderate slope for large SNs. However, we point out that this slope mostly results from the rightmost two points in figure \ref{Fig-DailyWvsSN}. Those two means are based only on a few daily pairs, contrary to lower SNs, which is reflected by the much larger errors for those points. Taking into account the actual errors for each bin in our regression, we find a lower value for the slope than \citep{Svalgaard_2015a} and we can also conclude that a constant W factor is compatible with the data. The linear regression leads to: 
\begin{equation}
W= 1.162 (\pm 0.016) + 0.00015 (\pm 0.00015) R_i^*       \label{EQ-Wreg>50}
\end{equation}
while a simple average for all $R_i^* > 50$ gives a mean value $W= 1.177 \pm 0.005$. We thus conclude that the inflation factor W remains close to 1.177 for all SN values larger than 50, i.e. over the main active phase of past solar cycles back to 1947, except near the cycle minima. Moreover, when adopting the slight upward dependency given by our regression, we can see that even for the largest cycles peaking at $S_N = 280$ (cycle 19), the W factor just reaches 1.2 but cannot exceed it. We note that this saturation of the W factor, and thus its low dependency to $R_i^*$ above 50, strongly reduces the possible error made when extrapolating the W factor for the largest observed SN values.

\begin{figure} 
	\centerline{\includegraphics[width=1.0\textwidth,clip=true,trim= 0 0 10 0,clip=true]{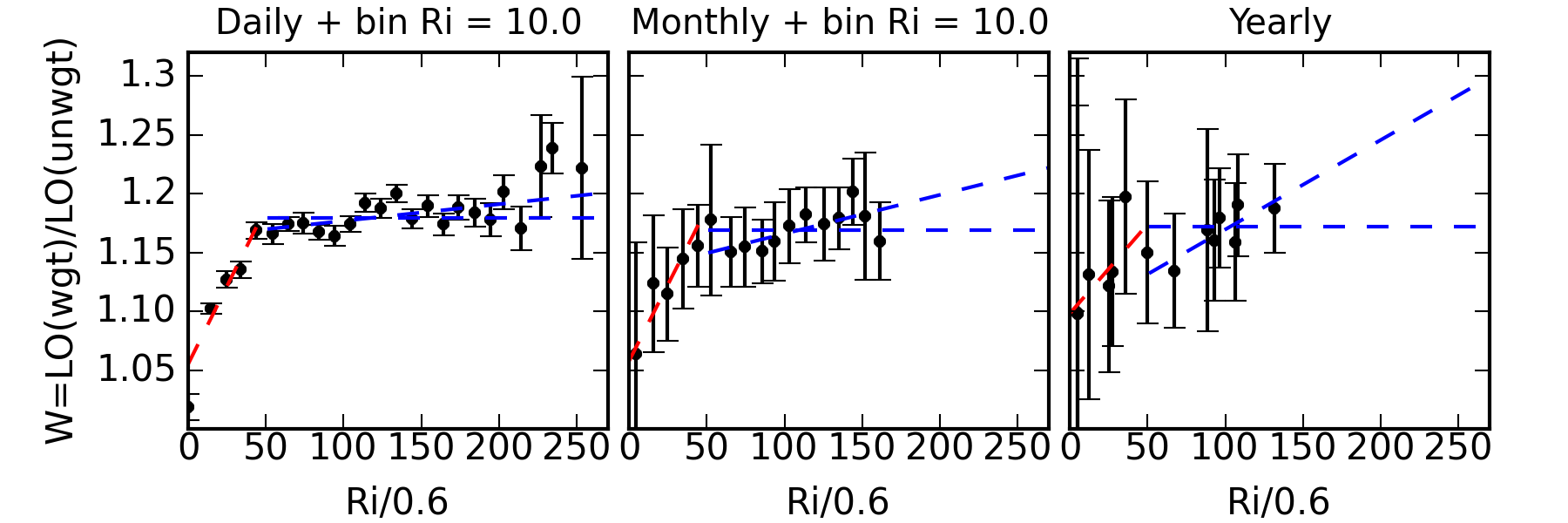}}
	\caption{Variation of the sunspot weigthing factor W as a function of $R_i^* = R_i / 0.6$ derived after averaging daily values over three different durations: 1 day, one month and one year. For each sub-plot, we include the same three fits as in figure \ref{Fig-DailyWvsSN}. The saturation at high $R_i^*$ progressively disappears for longer averaging durations, finally hiding the presence of a maximum limit to W.} 
	\label{Fig-W3timeAverages}
\end{figure}

The existence of this asymptotic value of the W factor is thus in contradiction with the early fit on annual means (Equation \ref{EQ-WregSvalg}), which requires a more careful interpretation. For this purpose, we repeated the determination of the $W(R_i)$ relation for different time averaging intervals of the base daily pairs of values. We illustrate the results in figure \ref{Fig-W3timeAverages} for three averaging durations: 1 day, 1 month and 1 year. We observe that as the averaging interval increases, the sharpness of the transition at intermediate $R_i^*$ decreases: the initial rise becomes more progressive, while the upper plateau takes a slope. The transition point is also shifting progressively to higher $R_i$ values. Those transformations can be explained by the mixing of days with different activity levels (i.e. wider range of SN values) when averaging over a long duration. This produces a smearing of SN values and thus of W factors over that range, with the mean W factor always getting lower. This leads to a larger uncertainty in the resulting W value and to a linearization of the $W(R_i)$ function. Moreover, it turns out that for yearly means, the asymptotic value is only reached for the highest SN values included in the observations, thus just missing the inflexion to a plateau at high $R_i^*$. Our results thus show that time-averaging of the W factor leads to misleading results and demonstrate that extrapolating the initial linear fit on yearly means is fully incorrect. 

In order to derive the new $S_N$ values, we thus applied the W correction factor given by the expression \ref{EQ-Wreg<50} for $R_i^* \leq 50$ and by the constant factor 1.177 for $R_i^* > 50$ to the original daily $R_i$ values. The monthly mean and yearly mean SN values were then obtained by averaging those corrected daily values. This remains an approximate correction, as the weighting factor actually varies for each individual active region observed on one specific day. However, this hardly affects the correction for high activity levels and thus cycle maxima. The correction factor is more approximate at low activity, but as the factor then drops closer to 1 and the SN values are low, the resulting absolute difference in SN is actually smaller than for high $S_N$ values.

\section{The ``Schwabe -- Wolf'' correction (1849 -- 1863)} \label{S-WolfCorr} 
Before 1947, the production of the SN was only based on the standard definition of the Wolf Number. At this stage, we thus kept the original numbers as provided by the successive prime observers of the Z{\"u}rich Observatory, including the historical reconstruction made by Wolf before 1849. This early part will definitely deserve a critical revision, in particular the annual means over 1700 -- 1750, for which the data are scarce. Among those possible defects, a particularly critical transition occurs in 1849, between the compiled historical SNs and Wolf's own systematics counts. Over this interval, H.\,Schwabe was the primary long-duration observer available to Wolf, providing a continuous series of observation from 1826 to 1867, thus overlapping Wolf's observations \citep{Arlt_2011, Arlt_etal_2013}. Schwabe's observations thus played a key role to ensure the scale homogeneity on both sides of the 1849 transition. 

\subsection{Confirmed scale jump in 1849}

It turns out that by analyzing the 1826 -- 1867 Schwabe interval, \citet{Leussu_etal_2013} found a large downward jump of $\sim 20 \%$ in the scale of the SN when directly compared to Schwabe's raw Wolf Numbers. This leads them to conclude that the entire early part of the SN series is overestimated compared to the recent SN and that all values before 1849 should be lowered by 20 \%. 

\begin{figure} 
	\centerline{\includegraphics[width=1.0\textwidth,clip=true,trim= 5 0 5 0,clip=true]{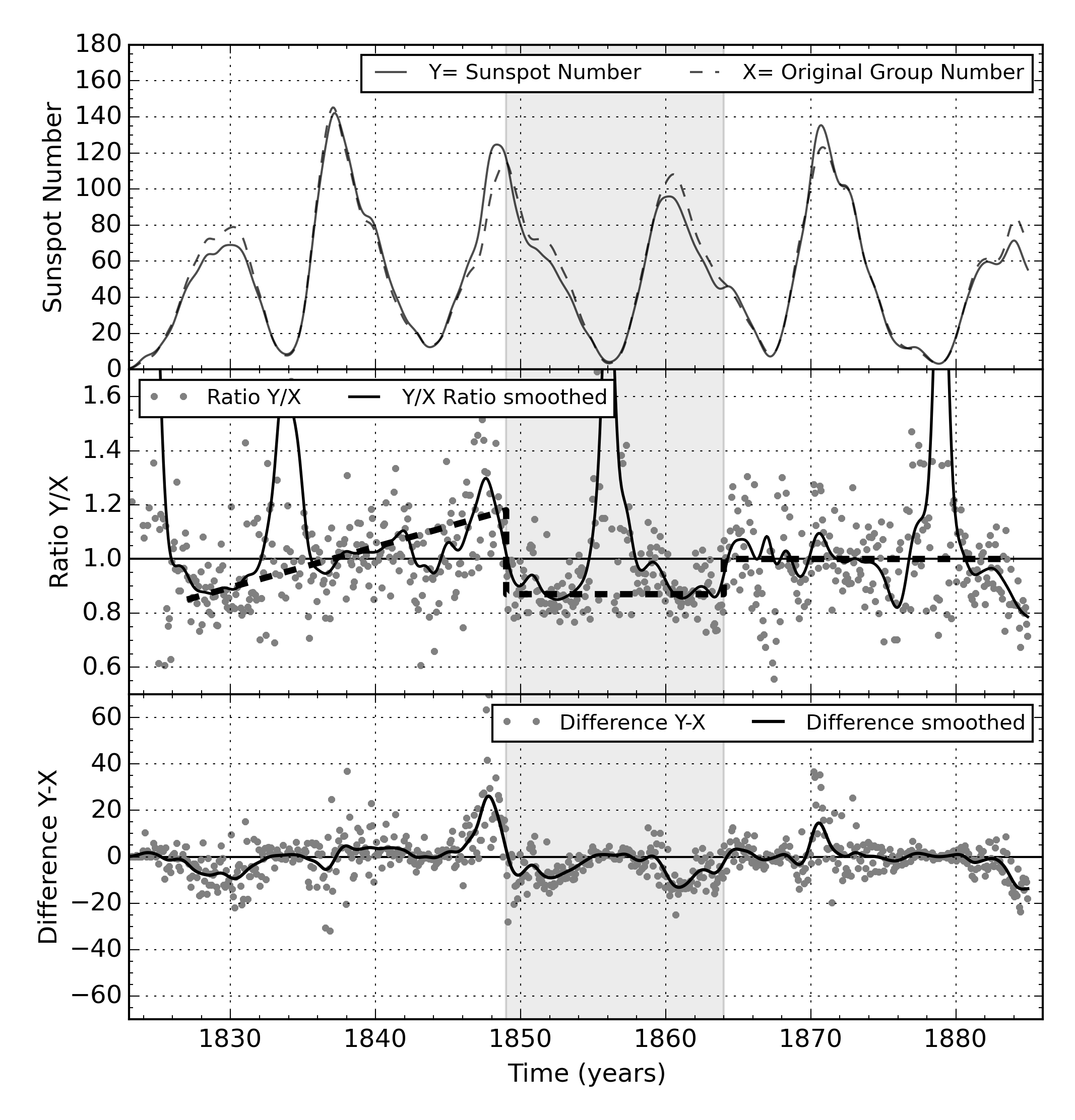}}
	\caption{Comparison of the original SN series (upper panel, solid line) with the original GN series (dashed line) over the interval 1823 -- 1885. For clarity, the monthly data were smoothed by a 12-month Gaussian function. The corresponding SN/GN ratio and SN-GN difference are plotted in the middle and lower panels, with dots corresponding to the un-smoothed monthly mean values. The broken dashed line in the middle panel marks the main trends and jumps present in the data and described in the main text.} 
	\label{Fig-SNGNratio1849old}
\end{figure}

\begin{figure} 
    \centerline{\includegraphics[width=1.0\textwidth,clip=true,trim= 5 0 5 0,clip=true]{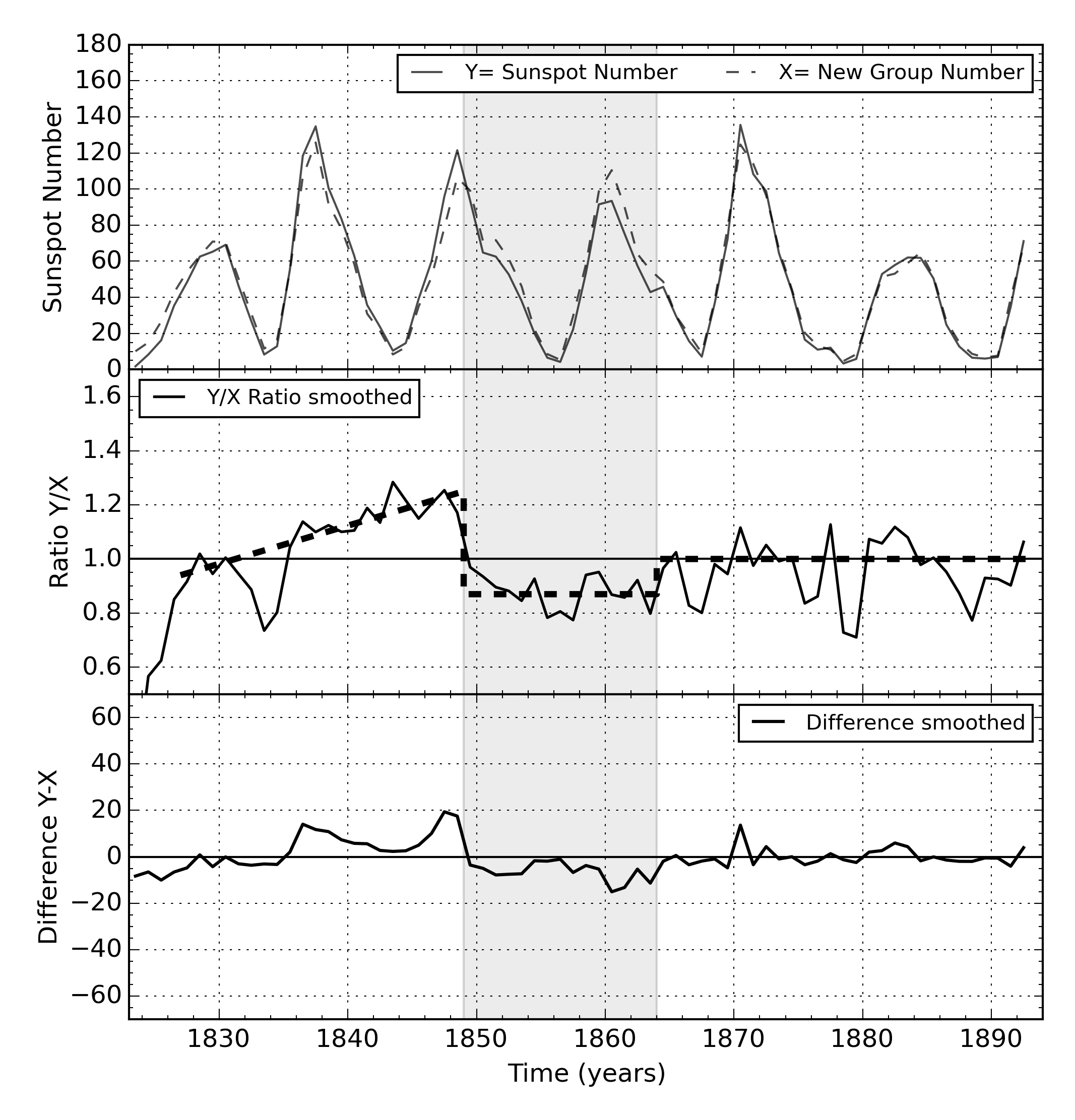}}
	\caption{Comparison of the original SN series (upper panel, solid line) with the new ``backbone'' GN series (dashed line) over the interval 1823 -- 1893. In this case, only yearly means are available but the GN series does not suffer from the inhomogeneity affecting the original GN after 1885. The corresponding SN/GN ratio and SN-GN difference are plotted in the middle and lower panels. Like in figure \ref{Fig-SNGNratio1849old}, the broken dashed line in the middle panel marks the main trends and jumps present in the data and described in the main text. } 
	\label{Fig-SNGNratio1849new}
\end{figure}

In order to check the existence and amplitude of this jump, we compared the original SN series with both the original and new ``backbone'' GN over a wider time interval 1826 -- 1880. Over this interval, both GN series are in good agreement, indicating that a major inhomogeneity is unlikely. We exclude the GN data before 1826, when the combination of a small number of observers and very low SN values in the Dalton minimum leads to a low accuracy and likewise, we exclude the large ``Greenwich'' trend found in the original GN after 1885 (see section \ref{SS-WaldContradic}). The series and corresponding ratios, shown in figures \ref{Fig-SNGNratio1849old} and \ref{Fig-SNGNratio1849new}, confirm the sharp downward transition in 1849. The amplitude of the jump is about 21\%, i.e. very close to the \citet{Leussu_etal_2013} value. 

Here in addition, we  observe that the jump is preceded by an upward trend in the ratio, which is more marked for the original GN ($+5\%$ between 1826 and 1849) than for the ``backbone'' GN ($+2.3\%$). In fact, this upward trend was also found in the study by\citet{Leussu_etal_2013} . As it even appears in the ratio between Schwabe's own sunspot counts and group counts, it indicates that Schwabe counted progressively more spots in the early part of his observations, mainly from 1836 to 1840. As indicated by \citet{Arlt_2011}, this inhomogeneity reflects the improving accuracy in Schwabe's early drawings, which include more sunspot details after 1830. Therefore, when considering all values before 1826, which were tied to the early Schwabe observations, this trend must be taken into account as it partly reduces the actual scale jump of 1849 relative to the early part of Schwabe's time series.

\subsection{Scale jump in 1864}

More importantly, our $S_N/G_N$ ratio shows another jump occurring in 1864, this time upwards, with fairly constant values from 1849 to 1863 as well as after 1863. Taking the ratio of the means over the intervals 1849 -- 1863 and 1864 -- 1885, we find a difference of $14\% \pm 1.8\%$ using either GN series. This scale jump is thus independent of the chosen GN reconstruction. It was obviously missed in the analysis by \citet{Leussu_etal_2013}, as they limited their study to the interval of the Schwabe series, which ends just after the second jump in 1867. 

Now, regarding the long-term homogeneity of the SN scale, this second jump largely compensates the -20\% jump of 1849. Together with the initial Schwabe trend, we end up with essentially identical scales in 1826 and after 1863, indicating that Wolf has properly scaled the early part of the series. When using the original GN as reference, we actually find that the SN in 1826 is $-7\% \pm 3.5 \%$ lower than the post-1864 values. With the ``backbone'' GN, the scale difference is smaller and barely significant, with the SN in 1826 $+4.2\% \pm 2.5\%$ higher. 

When looking for an objective reason for the 1863 jump, we noticed the coincidence of different historical elements. While Wolf defined his number and started his systematic observations in 1849 \citep{Wolf_1856_I}, he only published its full expression including the k personal coefficient in 1861 \citep{Wolf_1861_XII}. Over this early interval, he filled in the daily gaps in his observations using Schwabe's numbers, but he apparently only realized progressively that their respective counts were systematically different. Then, starting in 1861, several changes took place in close succession. 

\begin{figure} 
	\centerline{\includegraphics[width=1.0\textwidth,clip=true,trim= 15 0 5 0,clip=true]{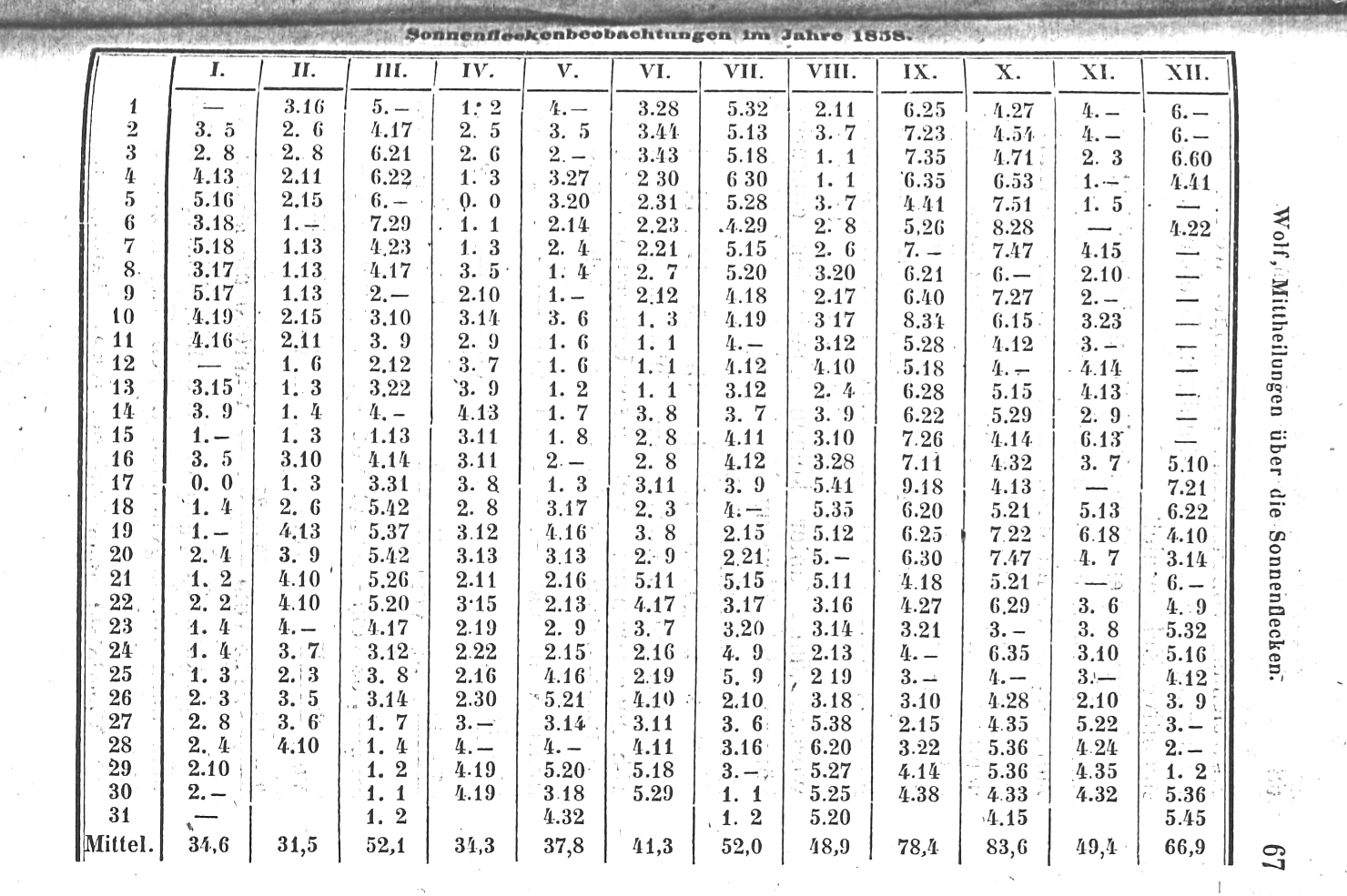}}
	\centerline{\includegraphics[width=1.0\textwidth,clip=true,trim= 15 0 5 0,clip=true]{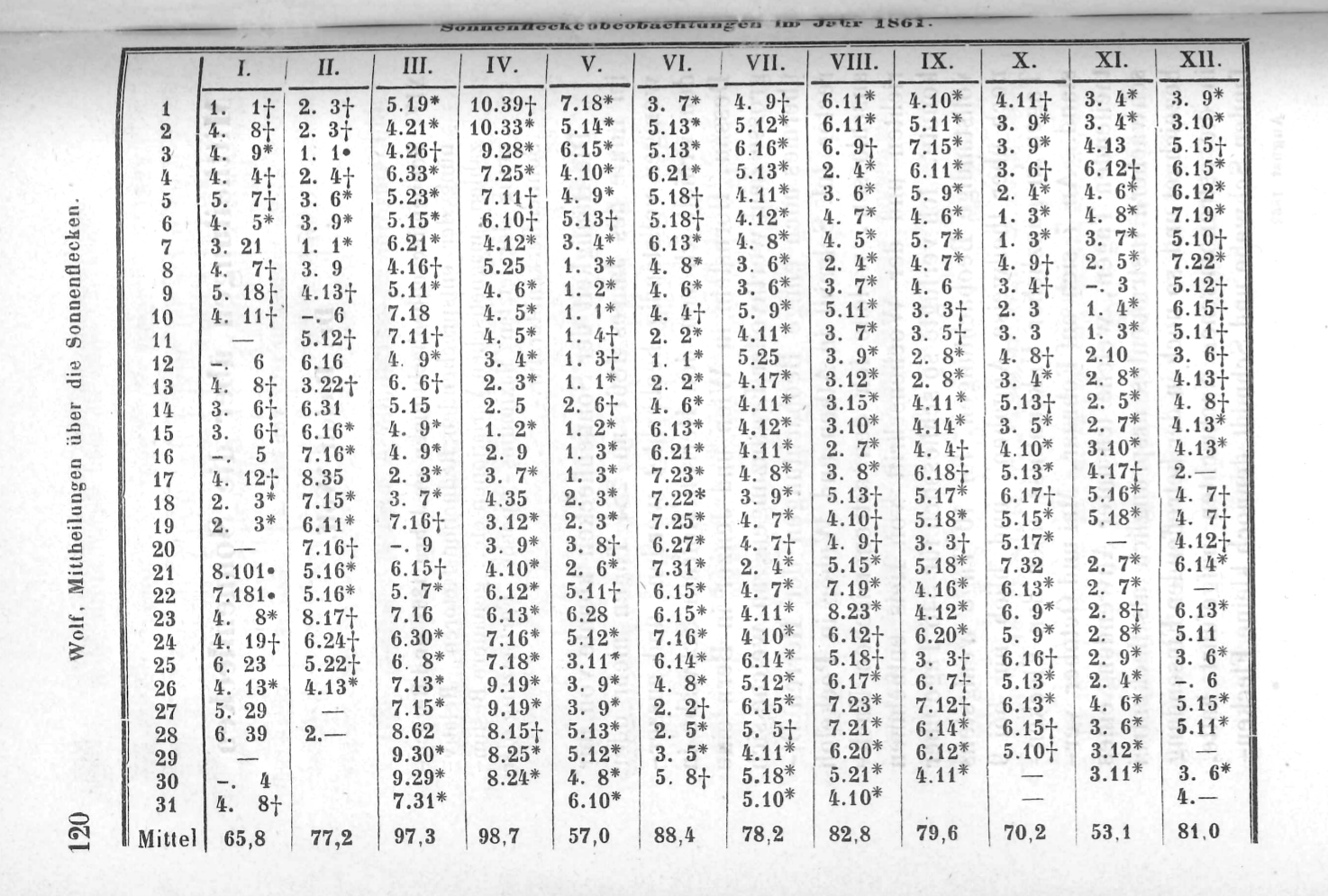}}
	\caption{Reproduction of the original yearly tables published by R. Wolf for the year 1858 (top) and 1861(bottom) \citep{Wolf_1859_VIII,Wolf_1862_XIV}. It shows the appearance of markers identifying the different observers, which were absent before 1861. Note also that most days in 1861 are marked by a * identifying Wolf using the small ``Pariser'' refractor. Only a few days are without marker, corresponding to Wolf observing with the standard 83\,mm refractor.)} 
	\label{Fig-WolfTables}
\end{figure}

In 1861, Wolf started to include additional auxiliary observers and to determine their $k$ personal normalization coefficients \citep{Wolf_1861_XII}. Until 1861, the daily observer was not identified in the published tables, making it very difficult to distinguish between Wolf's own counts and alternate counts from Schwabe. Only starting with the 1861 table \citep{Wolf_1862_XIV}, did Wolf include different symbols in the tables at the end of each line, to mark different observers (Figure \ref{Fig-WolfTables}). In 1864, for the annual data from 1863, Wolf started to produce a new table of daily relative sunspot numbers based on an average of all observers, next to the base table giving the number of groups and sunspots, thus further developing the normalization of all observers \citep{Wolf_1864_XVI}. However, it is probably in 1865 that the most important transition took place, for the observations of 1864. On that year, a first assistant joined Wolf at the Z{\"u}rich Observatory (Wielenmann) and started to observe sunspots with the standard 83\,mm refractor. Until then, Wolf contributed the bulk of the observations, but, as he was often traveling for official duties, he used a small portable ``Pariser'' telescope which led to significantly lower counts and required a 1.5 correction factor to adjust the Wolf number to the scale of the standard refractor \citep{Clette_etal_2014}. The observer markers in the 1864 table \citep{Wolf_1865_XVII} show that from July 1864 onwards, most of the daily counts were obtained with the standard refractor, probably mostly by Wielenmann, by contrast with the earlier Wolf-only period. This thus marks a sharp change to the main reference counts.

Therefore, we infer that the underestimate in the SN between 1849 and 1863 probably results from an imperfect combination of counts by Wolf and Schwabe and the dominant use of the small portable telescope until mid-1864. This is the most likely interpretation, based on the limited information at our disposal. As published tables for 1849 -- 1861 don't indicate which daily observations were made by Wolf or Schwabe and don't distinguish the observations made by Wolf either with the standard refractor or the small travel telescope, it is currently impossible to reconstruct the SN from raw observations for that period. Only the recovery of handwritten logbooks may help clarify entirely this anomaly. It turns out that Wolf's personal registers were recovered very recently and can now be analyzed (see \citet{Friedli_2015a}, in this issue). We found that Wolf himself was puzzled by an anomaly in the year 1864, wondering why the yearly mean for 1864 was higher than in 1863, marking sharp break in the normal decline of cycle 10 \citep{Wolf_1865_XVII, Wolf_1866_XXI}.  

The correction finally adopted for this time interval consists in multiplying the original SN values by a constant factor 1/1.14 from 1849 to 1863, thus with sharp limits at both ends. The latter seem to reasonably reflect the actual sharp changes in the method and data composition, which were implemented on specific dates. The first one corresponds to the start of systematic observations by Wolf on January, 1, 1849, while the second one matches the arrival of a new key observer using the standard refractor in Z{\"u}rich in 1864. The local break in the decline of cycle 10, noticed by Wolf, largely vanishes after correction, leading to a normal continuous evolution for this cycle. This anomalous feature thus seems to be entirely attributable to this artificial discontinuity. 

Overall, this correction remains local without impact over the long-term scale of the SN, which was thus left unchanged over the intervals 1700 -- 1849 and 1865 -- 1946. At this stage, we did not correct for Schwabe's trend affecting the period 1826 -- 1849 as this is a smaller correction and the existing GN series don't fully agree on this period, preventing to derive a correction with an accuracy better than the amplitude of this defect. Our analysis also suggests that this local trend has no significant long-range influence on the overall uniformity of the SN series. 

We must stress that although we use here a comparison with the GN series, it only involves a short period around each jump (excluding time intervals over which major corrections to the GN were needed). We consider only relative variations after matching the average scales of SN and GN over the stable interval 1863 -- 1885. Therefore, the absolute long-term scale of the GN does not play any role in this comparison. Luckily, thanks to the sharpness of the jumps, we could work here on rather short intervals. We also know that such sharp jumps in 1849 and 1864 are specific to the construction of the SN and thus don't coincide with any equivalent transition in the completely different construction of the old and new GN series.   

\section{Discussion} \label{S-Discussion} 
\begin{figure} 
	\centerline{\includegraphics[width=1.0\textwidth,clip=true,trim= 5 0 5 0,clip=true]{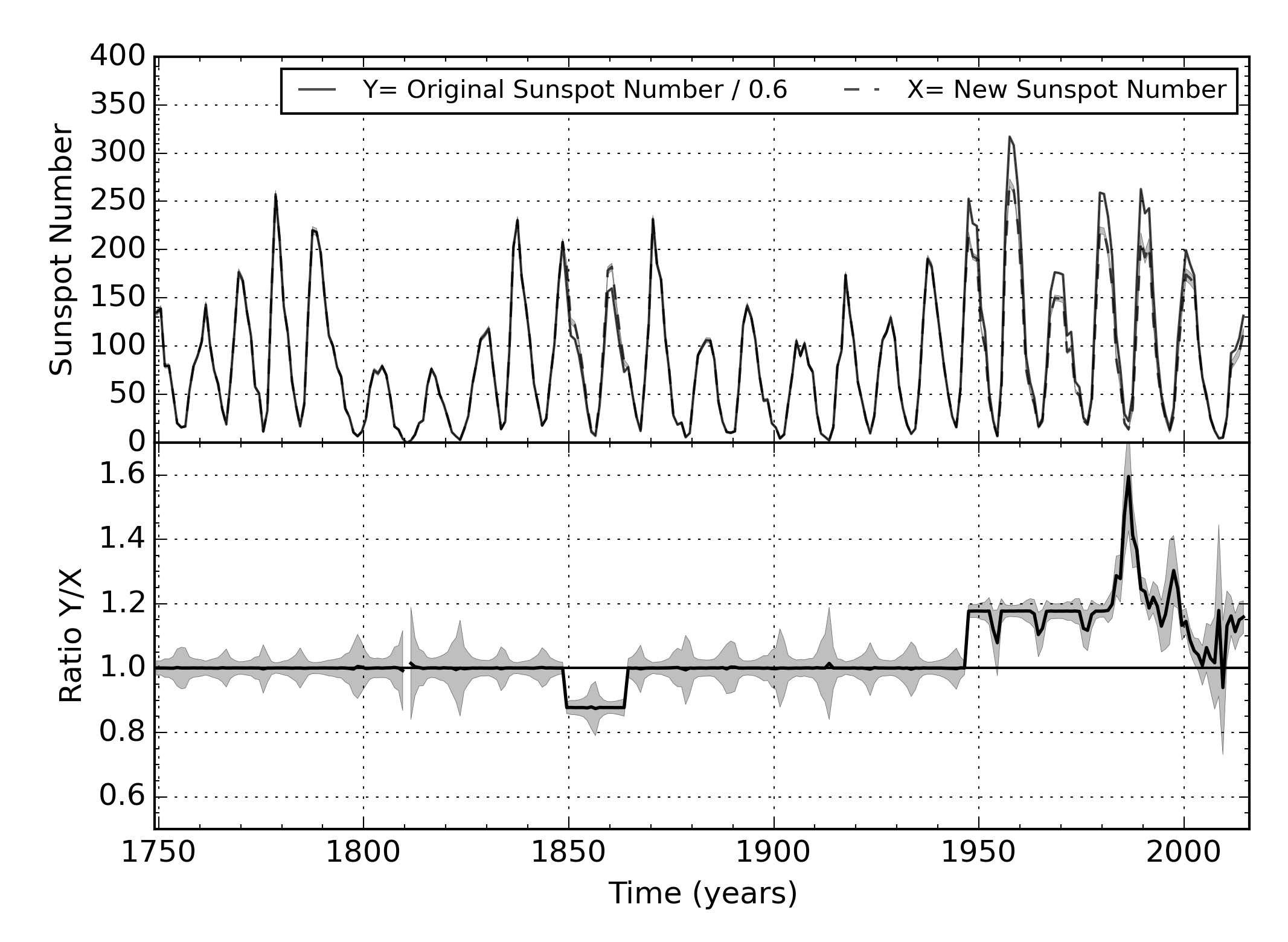}}
	\caption{Comparison of the original SN (upper panel, solid line) and new SN (dashed line). For readability, we show here the 12-month Gaussian-smoothed monthly means. The original SN was divided by the 0.6 conventional Z\"urich factor, which gives a ratio of unity over most of the early part of the series up to 1980 (lower panel). The deviations from unity in the ratio correspond to the three corrections applied to sections of the original series for producing the new revised version. The grey shading indicates the standard error on the ratio.} 
	\label{Fig-SNallCorr}
\end{figure}

After combining all the above corrections, we can compare the new and original SN series (Figure \ref{Fig-SNallCorr}), where the monthly mean ratios closely correspond to the applied corrections. As the two main corrections modify the relative amplitude of the SN over the last decades relative to the earlier values, the correction can have implications on various topics and studies resting on this long-term solar activity record. Here, we consider just two immediate and illustrative comparisons. As a key initial issue was to understand and possibly resolve the discrepancies between the SN and GN series, we first re-do this global SN and GN comparison.

\subsection{Better agreement between the Sunspot Number and Group Number}

\begin{figure} 
	\centerline{\includegraphics[width=1.0\textwidth,clip=true,trim= 5 0 5 0,clip=true]{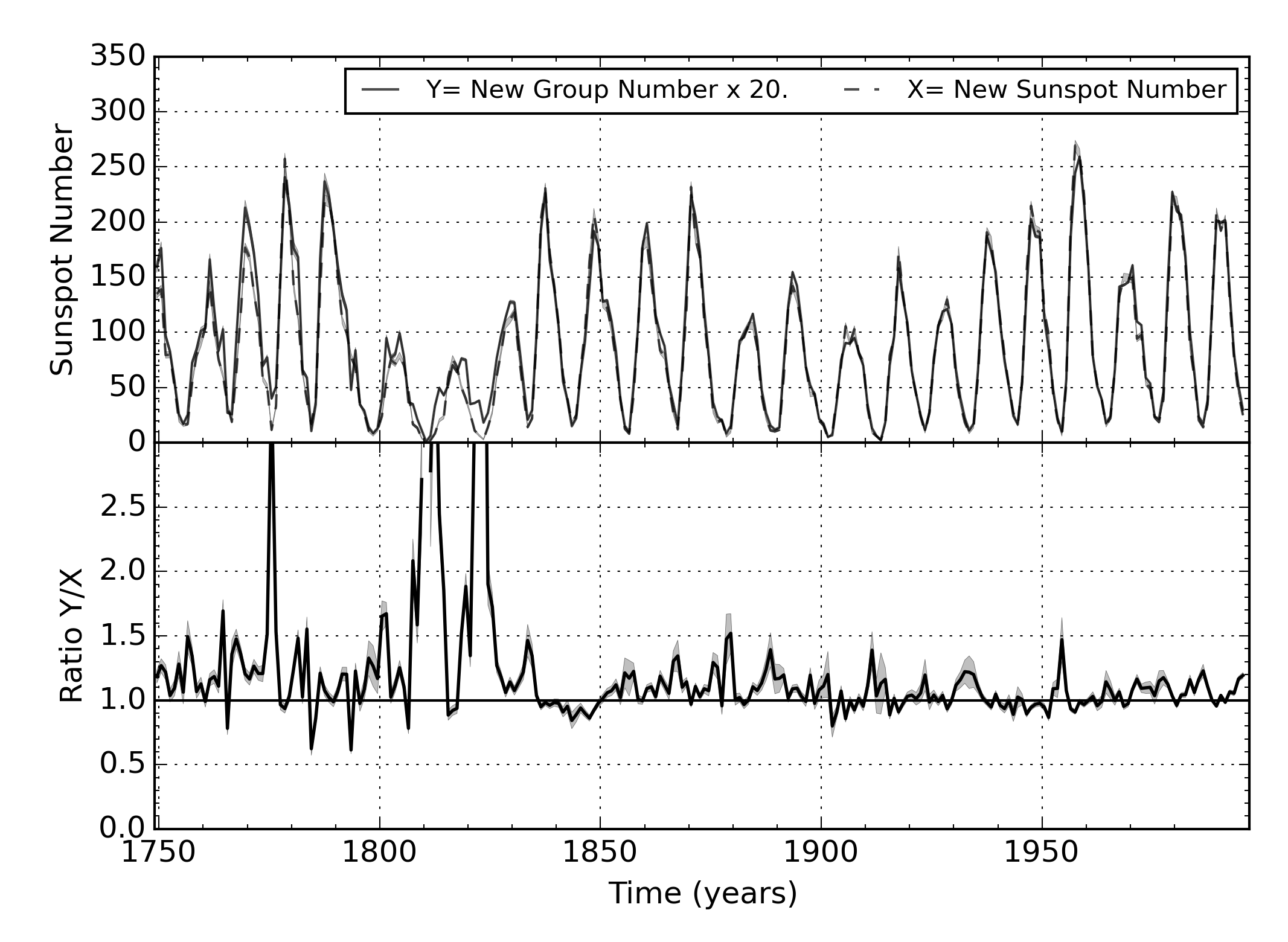}}
	\caption{Comparison of the yearly means of the new SN (upper panel, dashed line) with the new ``backbone'' GN (solid line). The GN was multiplied by 20 to match the scale of the SN over the $20^{th}$ century. The lower panel shows the GN/SN ratio, with the grey shading giving an estimate of the uncertainty based on the errors provided for the ``backbone'' GN \citep{Svalgaard-Schatten_2015}. The large persistent deviations present in the original comparison \citep[Fig. 1]{Clette_etal_2014} have now largely disappeared. The larger errors in the early data before Schwabe (1826) produce larger random deviations of the ratio in the early part. Note however that the largest peaks always correspond to solar cycle minima, and thus only reflect small absolute differences between the series when the SN values are close to 0.} 
	\label{Fig-compSNGN}
\end{figure}

The new ``backbone'' GN \citep{Svalgaard-Schatten_2015} consists in a full reconstruction using a new approach that avoids the identified causes of inhomogeneities in the original GN series: primarily the 1885 -- 1915 40\% drift associated with the use of the Greenwich photoheliographic catalog and also a $\approx 10\%$ jump after 1976 when Greenwich data were extended by a few alternate observers \citep{Clette_etal_2014}. The main 1885 correction is also confirmed and interpreted by \citet[this volume]{Cliver-Ling_2015}. Therefore, in figure \ref{Fig-compSNGN}, we show a comparison between this new GN series and our current SN over the common 1749 -- 2015 time interval. For now, we can only compare yearly means, as only yearly means are available for this GN.

Compared to the original series \citep[Fig. 1]{Clette_etal_2014}, we find a much better agreement, with essentially a constant ratio from the end of the Dalton minimum (1826) to the present. It thus confirms that the main corrections applied over that period significantly reduced or eliminated actual defects affecting either of the original series. Note that the average ratio $S_N$/$G_N$ is now about 20., instead of the original 12.08 factor used by \citet{Hoyt-Schatten_1998a}. This is explained by the use of the Wolfer period (1877 -- 1926) as scale reference in both series, which results from the elimination of the 0.6 conventional factor in the case of the SN. However, although no long enduring trend remains, we observe that local deviations are still present and locally significant, given the estimated uncertainties.
We note that part of those deviations show a solar cycle modulation, and are thus probably associated with the non-linear relation between the SN and GN, as established by \citet[Paper 1 in this volume]{Clette_etal_2015a} for the recent decades.

Before 1826 (start of observations by H. Schwabe), the ratio shows much larger random variations, reflecting the larger uncertainties due to a combination of three main factors:
\begin{itemize}
	\item the low SN numbers during the weak cycles marking the Dalton minimum
	\item the lower number of observers and observations
	\item the cruder optical instruments available before the $19^{th}$ century
\end{itemize}
Therefore, the largest deviations from unity in the ratio mostly occur near cycle minima and actually correspond to small absolute differences in the values. Still, by determining the average ratio between the two series over the 1749 -- 1800 interval using multiple methods as in \citet[Paper 1]{Clette_etal_2015a}, we find that the ``backbone'' GN is higher than the SN, on average by about 20\%.

Overall, we thus conclude that both series definitely mark a large improvement in accuracy. This is particularly true for all values after 1826. The earlier part will clearly require further analyses in coming years. For this new version of the SN, the early historical numbers compiled by R. Wolf were not revised. Therefore, the main progresses can be expected from the recovery and verification of observations in the scarcely covered interval between Staudacher and Schwabe (1795 -- 1824) and also in the first half of the $18^{th}$ century. The return of the solar cycle after the end of the Maunder Minimum thus still involves large uncertainties at this stage, although the new ``backbone'' GN series indicates stronger cycles in the early $18^{th}$ century and thus a better agreement with the SN series, as created by Wolf (Figure \ref{Fig-comparTrends}).
 
\subsection{No secular trend over the past 250 years}
\begin{figure} 
	\centerline{\includegraphics[width=1.0\textwidth,clip=true,trim= 5 0 5 0,clip=true]{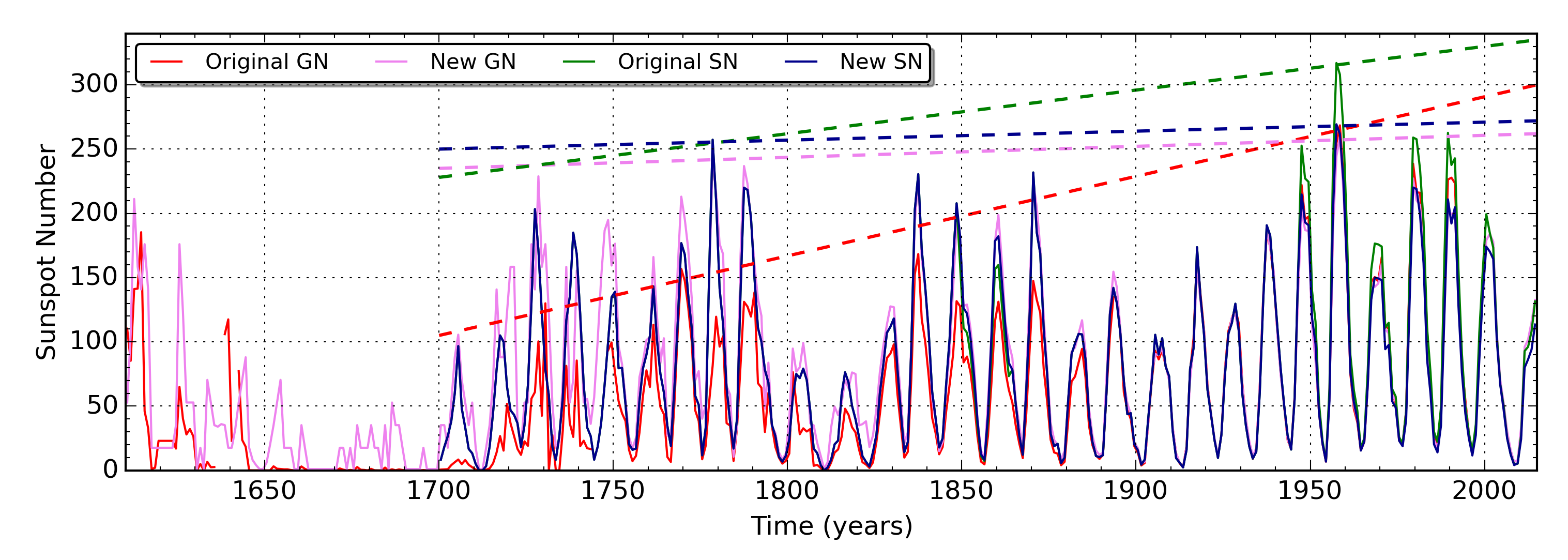}}
	\caption{Comparison of secular trends before and after correcting the SN and GN series: original SN (green), original GN (red), corrected SN described in this paper (blue) and new ``backbone'' GN (purple). As a visual guide, the dashed lines give the overall trend in all four series (with matching color), by connecting the highest maxima of the $18^{th}$ (cycle 3) and in the $20^{th}$ centuries (cycle 19). Both new series show the same absence of a rising trend over the last 3 centuries. } 
	\label{Fig-comparTrends}
\end{figure}

While both the original SN and GN series showed higher solar cycle amplitudes in the $20^{th}$ century compared to the preceding centuries, the original SN already indicated a lower trend ($\approx 20\% / century$) than the GN series ($\approx 40\% / century$). After the corrections described above, we find that the scale of recent cycles was in fact overestimated in the SN compared to all earlier cycles. An equivalent conclusion was found for the GN but due to a totally different cause (trend in the Greenwich photographic data) and with an even larger correction.  Now comparing the trend in the amplitudes of the largest cycles over several centuries (Figure \ref{Fig-comparTrends}), we can see that the new SN displays a very weak trend almost identical to the new GN. Here again, we must stress that this agreement results from completely independent corrections to both series.

Therefore, our present results confirm that the earlier strong upward trend characterizing the original GN was an artifact and definitely not a solar property. In addition, as the corrections further reduce the smaller trend present in the original SN series, the residual difference between cycle maxima in the $20^{th}$ century versus earlier ones falls below the remaining uncertainties, thus a few percents, i.e. at least an order of magnitude lower than indicated by the original uncorrected series. This thus strongly questions the notion of a modern Grand Maximum \citep{Solanki_etal_2004_Nature, Abreu_etal_2008_GeoRL, Lockwood_etal_2009_ApJ700, Usoskin_etal_2014_AAp}. 

Next to the upward trend in the original GN series, this ``Grand Maximum'' concept rests primarily on an unusually high maximum in the mid-$20^{th}$ century found in time series of cosmogenic isotopes and attributed to a solar origin \citep{Usoskin_etal_2002_JGRA107, Usoskin_2013_LRSP}. We note however that all studies do not agree on this interpretation: a recent one derives rather constant levels over the last centuries \citep{Muscheler-Heikkila_2011}. For calibrating those time series, the solar influence is translated into the modulation potential $\phi$, which results from the elimination and compensation of many factors unrelated to the Sun: deposition processes, evolution of the Earth magnetic field, etc.. 

\begin{figure} 
	\centerline{\includegraphics[width=0.85\textwidth,clip=true,trim= 0 0 2 0,clip=true]{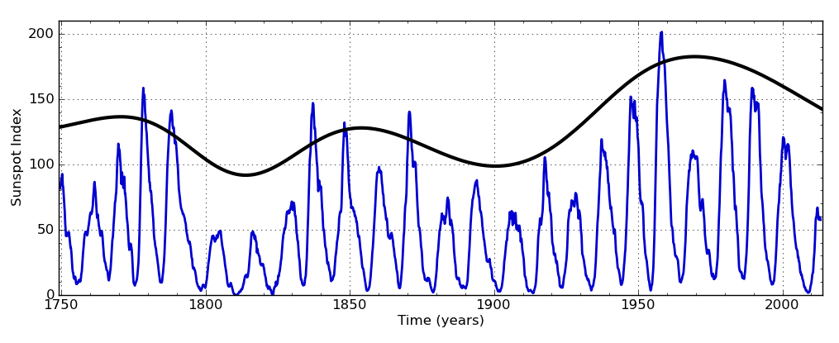}}
	\centerline{\includegraphics[width=0.85\textwidth,clip=true,trim= 0 0 2 0,clip=true]{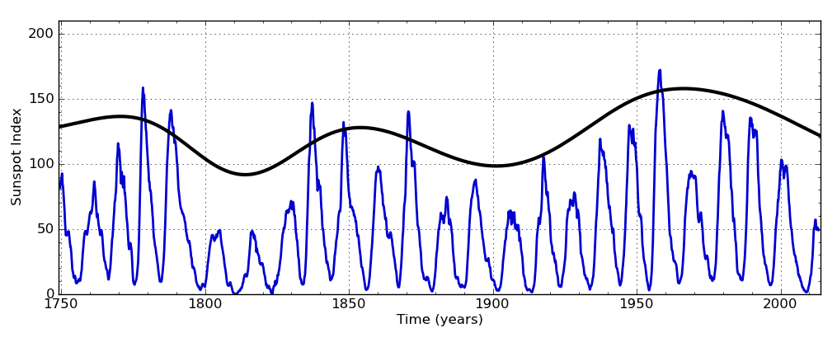}}
	\caption{Effect of 22-year Gaussian smoothing on the secular variations of the original SN series (top) and the new revised SN (bottom). The blue curves show the 12-month Gaussian smoothed monthly means, while the thick black curves are the 22-year smoothed values (here multiplied by a factor 3 to better outline the overall envelope of the cycle amplitudes). For the original SN series, the last maximum of the 22-year smoothed SN is 31\,\% higher than earlier peaks of past centuries (top). Although cycles amplitudes remain within a constant range for the new SN, the 22-year smoothing still leads to a slightly higher peak, by 17\,\%, in the $20^{th}$ century (bottom).} 
	\label{Fig-smoothTrends}
\end{figure}

In addition, we note that given the limited time resolution available in the cosmogenic isotope record, this potential $\phi$ is typically averaged over durations longer than the solar cycle, up to 50 years \citep{Muscheler-Heikkila_2011}. In order to simulate this, we applied a 22-year running mean to the original and corrected SN series (Fig. \ref{Fig-smoothTrends}). We find that while the difference of peak amplitudes of the cycles between the $19^{th}$ and $20^{th}$ centuries drops from 30\% to almost 0, the running mean still shows a difference using the new SN series, though reduced by half. Therefore, we argue that a higher solar signal in the cosmogenic record over the $20^{th}$ is not necessarily in contradiction with equal cycle amplitudes over the past centuries. Through the time averaging, the higher solar modulation potential $\phi$ only reflects the longer sequence of strong cycles in the $20^{th}$ century compared to similar episodes in the previous centuries. This would mean that the notion of Grand Maximum does not imply exceptionally high cycles amplitudes as previously thought. So, to the extent that this notion remains valid for the cosmogenic record, the definition of a Grand Maximum needs to be reconsidered. 

\section{Conclusions} \label{S-Conclusions} 
For each correction described above, we had to make a choice of the best estimate based on current available results and data. This leads to the current new instance of the SN series, which represents a major improvement versus the original SN series. Even if more accurate determinations can be expected and validated in the future, we think that the release of this new version is fully timely. Indeed, the three main corrections applied to the series are supported by several parallel analyses. Moreover, the corrections reach up to 20\%, which can definitely affect the results and conclusions of analyses based on the SN. Therefore, it is essential and immediately beneficial to abandon the original flawed data for a largely better version. Indeed, we definitely reduce the inhomogeneities by a factor 4 or more, whatever the choice of the optimal correction, given the maximum range of possible values left at this stage.

In fact, the release of this new version of the SN is not simply delivering a new data set. More importantly, it marks a fundamental transition between the earlier unalterable and unquestioned data series to a genuine measurement series, like any other physical data series. Like for any other measurement, it is natural to revise it as new data sources become available and new analysis methods become available. This also means that we must prepare for future upgrades of this unique data set and try to set up a structured process to implement and document future changes. This is why, together with the release of the new SN series by the WDC - SILSO, we implemented a new version tracking scheme:
\begin{itemize}
	\item Successive versions are identified by a unique version number. This number is included in the filename and directory names containing the data.
	\item Only one version, the latest one (highest version number) is distributed as the master series in the main SILSO data page. Only this version is maintained and extended by appending new SN values derived from observations from the SILSO network.
	\item All past versions remain accessible through a dedicated entry in the dedicated ``Archive'' section of the SILSO Web portal. Those past versions are mainly provided as reference, e.g. if a past published analysis needs to be reproduced in order to discriminate changes associated with the use of a more recent version of the SN from other factors.
	\item Each version is documented, with a description of the file contents and information about the corrections applied relative to the earlier versions and of course, with the release date. This information may consist in explanatory files stored with the data files, scientific publications describing the related analyses and if relevant, ancillary data files needed to derive the corrections.
\end{itemize}
The version number consists of two numbers N.n, where:
\begin{itemize}
	\item N is the main version number: it corresponds to any major modification of the SN series leading to systematic changes in the values over long time intervals.
	\item n is the sub-version number. It reflects either secondary changes that don't modify the primary $S_N$ values (updated error estimates, file formats) or minor local corrections to the time series itself, which no long-range influence on the global homogeneity of the series (correction or replacement of single values, e.g due to typos)
\end{itemize}
In this scheme, the original SN series is now labeled as version 1.0, while the current new release is numbered 2.0. In order to avoid confusion and an excessive adaptation work for our users, we consider that new version releases should be reasonably spaced in time. A minimum interval of one year seems to be appropriate. This must provide enough time to prepare and validate a new modification.
Regarding this validation, in the framework of the International Astronomical Union (IAU), the WDC -- SILSO initiated discussions for the creation of an advisory committee that will serve as a link between WDC -- SILSO and the scientific community and provide a more formal endorsement of new updated versions of this reference solar index. Next to solar physicists, this committee can include data processing experts as well as representatives of users. Its main roles would be:
\begin{itemize}
	\item The approval of new corrections to the SN series, proposed and prepared by WDC -- SILSO.
	\item Submitting suggestions and advices for new corrections and updates to the SN series, e.g. based on new published results
	\item Proposing additional data products or upgrades to services based on the SN data (data distribution, graphics, etc.)
	\item The designation of the SN series as one of the reference data sets in astronomy.
\end{itemize}

A last innovation coming with the new SN series was to provide the standard error for each SN value. Currently, errors are only provided for the recent SN values based on the WDC -- SILSO database starting in 1981. Over that period, as we can use data from a large sample of observers for each day (typically 20 to 40 observers), we derive the standard deviation of all observations. The standard error on the daily, monthly and yearly mean can then be derived. From now on, those errors will be routinely included in the new SN values appended each month based on new observations by the SILSO network. In the future, we plan to progressively include error estimates for the Z{\"u}rich values before 1981. However, in this case, a completely different approach will be needed as the current values are mostly based on a single observer at the Z\"urich station. In this case, we may instead estimate the absolute uncertainty range, which reflects the accuracy of the value instead of the statistical dispersion of simultaneous numbers (precision). This determination of errors is discussed more in depths in a companion paper in this issue \citet{Lefevre_etal_2015}. Here, we also initiate a radical change in the philosophy of the earlier SN production, where the values were given without any information about the uncertainties, with a lack of transparency on the computing method.

Overall, our goal is that the scientific community gets a more direct and transparent access to the new SN series and thereby, develops a new interest in this historical reconstruction of the past solar activity. The recent multiplication of publications presenting new investigations of the past sunspot record is an encouraging sign that through this epochal revision of this unique solar series, we rejuvenated it and injected new life in solar cycle research for many years in the future.  

\begin{acks}
	F.Clette and L. Lef\`{e}vre would like to acknowledge financial support from the Belgian Solar-Terrestrial Center of Excellence (STCE; \url{ http://www.stce.be}). Part of this work was developed in the framework of the SOLID project (EU $\rm 7^{th}$ Framework Program, SPACE collaborative projects, \url{http://projects.pmodwrc.ch/solid/}) and of the TOSCA project (ESSEM COST action ES1005 of the European Union; \url{http://lpcs2e.cnrs-orleans.fr/~ddwit/TOSCA/Home.html}).
\end{acks}

%
\bibliographystyle{spr-mp-sola}
\bibliography{Clette_SN2}  
%
%
%
%

\end{article} 
\end{document}